\documentclass{aa}
\usepackage{graphics,epsf,epsfig,rotating,times}
\usepackage{amsmath}
\usepackage{amssymb}
\begin{document}

\title{Temporal and spectral gamma-ray properties of \object{Mkn~421} above 250 GeV from CAT observations between 1996 and 2000}

\author{
F.~Piron \inst{3}\fnmsep\thanks{\emph{Present address:} GAM, CC 085 - B\^at. 11,
Univ. de Montpellier~II, Place Eug\`ene Bataillon, 34095 Montpellier Cedex 5, France} \and
A.~Djannati-Ata\"{\i} \inst{6} \and
M.~Punch \inst{6} \and
J.-P.~Tavernet \inst{4} \and
A.~Barrau \inst{4}\fnmsep\thanks{\emph{Present address:} Institut des Sciences
Nucl\'eaires, 53 avenue des Martyrs, 38026 Grenoble Cedex, France} \and
R.~Bazer-Bachi \inst{1} \and
L.-M.~Chounet \inst{3} \and\\
G.~Debiais \inst{2} \and
B.~Degrange \inst{3} \and
J.-P.~Dezalay \inst{1} \and
P.~Espigat \inst{6} \and
B.~Fabre \inst{2} \and
P.~Fleury \inst{3} \and
G.~Fontaine \inst{3} \and\\
P.~Goret \inst{7} \and
C.~Gouiffes \inst{7} \and
B.~Khelifi \inst{6} \and
I.~Malet \inst{1} \and
C.~Masterson \inst{6} \and
G.~Mohanty \inst{3}\fnmsep\thanks{\emph{Present address:} IGPP, University of California, Riverside, CA 92521, USA} \and\\
E.~Nuss \inst{2}\fnmsep$^\star$ \and
C.~Renault \inst{4}\fnmsep$^{\star\star}$ \and
M.~Rivoal \inst{4} \and
L.~Rob \inst{5} \and
S.~Vorobiov \inst{3}
}

\offprints{F. Piron, \email{piron@in2p3.fr}}

\institute{
Centre d'Etudes Spatiales des Rayonnements, Universit\'e Paul Sabatier, Toulouse, France (INSU/CNRS)
\and
Groupe de Physique Fondamentale, Universit\'e de Perpignan, France
\and
Laboratoire de Physique Nucl\'eaire des Hautes Energies,
Ecole Polytechnique, Palaiseau, France (IN2P3/CNRS)
\and
Laboratoire de Physique Nucl\'eaire et de Hautes Energies,
Universit\'es Paris VI/VII, France (IN2P3/CNRS)
\and
Nuclear Center, Charles University, Prague, Czech Republic
\and
Physique Corpuscul\-aire et Cosmologie, Coll\`ege de France et Universit\'e Paris VII, France (IN2P3/CNRS)
\and
Service d'Astrophysique, Centre d'Etudes de Saclay, France (CEA/DSM/DAPNIA)
}

\date{Received 10 April 2001 / Accepted 6 June 2001}

\abstract{
The $\gamma$-ray emission above $250\:\mathrm{GeV}$ from the BL~Lac object Markarian~421
was observed by the CAT Cherenkov imaging telescope between December, 1996, and June, 2000.
In 1998, the source produced a series of small flares, making it the second extragalactic source detected
by CAT. The time-averaged differential spectrum has been measured from $0.3$ to $5\:\mathrm{TeV}$, which is well fitted
with a power law: $\displaystyle\frac{\mathrm{d}\phi}{\mathrm{d}E}\varpropto E_{\mathrm{TeV}}^{-2.88\pm0.12^\mathrm{stat}\pm0.06^\mathrm{syst}}$.
In 2000, the source showed an unprecedented activity, with variability time-scales as short as one hour, as for
instance observed during the night between 4 and 5 February.
The 2000 time-averaged spectrum measured is compatible
with that of 1998, but some indication of a spectral curvature is found between $0.3$ and $5\:\mathrm{TeV}$.
The possibility of $\mathrm{TeV}$ spectral hardening during flares is also discussed, and the results are compared to those obtained on
the other $\mathrm{TeV}$ BL~Lac, Markarian~501.
}

\titlerunning{gamma-ray properties of \object{Mkn~421} above $250\:\mathrm{GeV}$ from CAT observations between 1996 and 2000}

\maketitle

%
\begin{figure*}[t]
\begin{center}
\hbox{
\epsfig{file=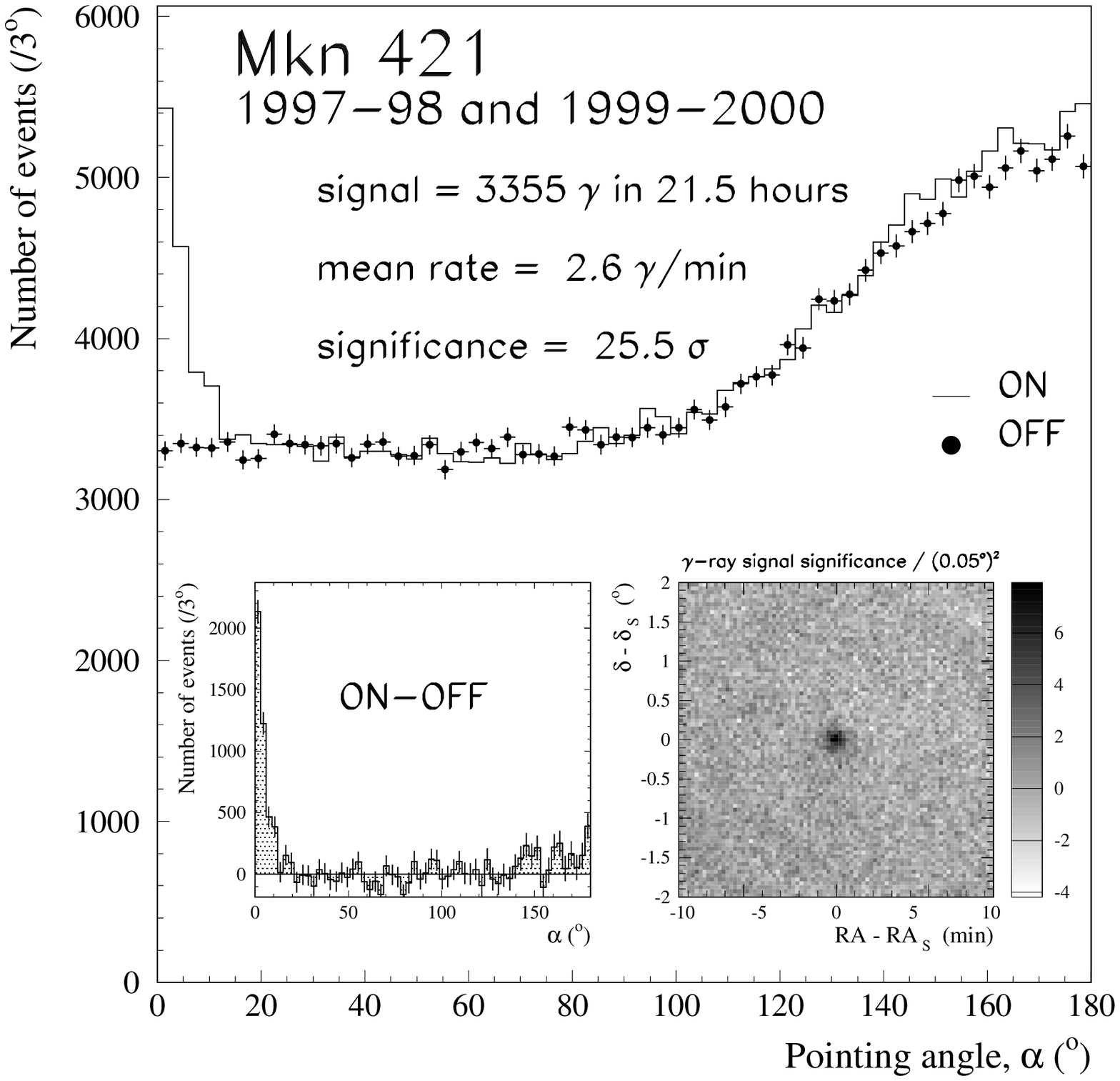,width=.49\linewidth,
,bbllx=25,bblly=156,bburx=531,bbury=649}
\epsfig{file=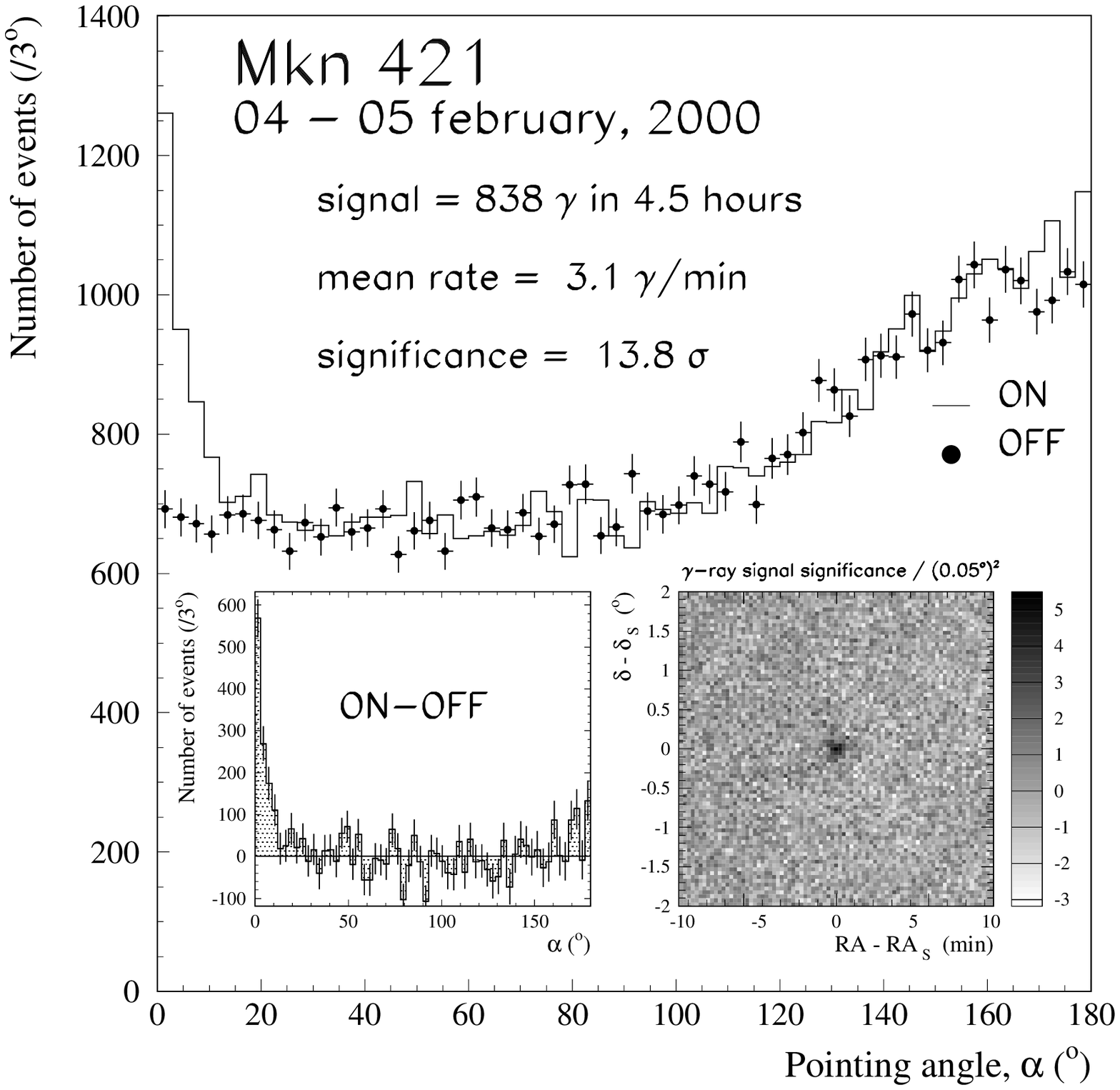,width=.49\linewidth,
,bbllx=25,bblly=156,bburx=531,bbury=649}
}
\caption{
The \object{Mkn~421} $\gamma$-ray signal seen with the CAT imaging telescope, for the most significant observations (signal significance greater
than $3\:\sigma$ in $\sim$$30\:\mathrm{min}$) taken in 1997--98 and 1999--2000 at zenith angle $\theta_{\mathrm z}$$<$$45\degr$ (left panel),
and for the highest flare, recorded between 4 and 5 February, 2000 (right panel).
In each panel, the main plot shows the pointing angle, $\alpha$, distribution for ON (solid line) and OFF data (points with error bars, same
duration as ON data), and
the bottom-left inset gives the ``ON--OFF'' distribution; events are selected by the
$Q_\mathrm{4}$, $Q_{\mathrm{tot}}$, and $\mathcal{P}(\chi^2)$ cuts (see text), and the values quoted in legend come from the additional cut $\alpha$$<$$6\degr$
(first two bins).
The bottom-right inset shows the significance map of event excess per bin of $0.05\degr$, obtained from the ``ON--OFF''
distribution of reconstructed angular origins (no cut on $\alpha$): the source lies at the centre, and the axes denote the relative right
ascension and declination with respect to the source coordinates ($RA_\mathrm{S}$ and $\delta_\mathrm{S}$).
}
\label{FigAlplot}
\end{center}
\end{figure*}
\section{Introduction}
Among Active Galactic Nuclei (AGNs), blazars are those radio-loud objects having a jet
pointing towards the observer, which has a relatively high bulk Lorentz factor, giving rise to a strong Doppler boosting of the observed
fluxes. Blazar emission is dominated by their jet power output, which is mainly non-thermal, extending over more than fifteen energy decades. At low
energies, their featureless optical continuum as well as their strong radio and optical polarization are due to synchrotron radiation of
the magnetized plasma jet. At high energies, blazar jets also show remarkable properties.
During the last ten years, the EGRET detector, operating above $30\:\mathrm{MeV}$ on board the Compton Gamma-Ray Observatory,
definitively opened the field of high-energy astrophysics by revealing that most extragalactic strong $\gamma$-ray emitters were
blazars (\cite{vonMontigny}; \cite{Hartman}). Their $\gamma$-radiation power often dominates their entire spectrum, and it must be produced in a
small enough region to account for rapid variability, already observed on time-scales less than one hour at $\mathrm{TeV}$ energies
(\cite{Gaidos}).
Although the origin of jets is still uncertain, the study of their $\gamma$-ray emission in blazars can
shed light on the nature and content of their plasma (e$^+$e$^-$ pairs or e$^-$p), and give new insight into
the high-energy particle acceleration and cooling processes occuring at the sub-parsec scale.
Ultimately this could give useful information on how jets take form and lead to a better understanding
of the energy extraction mechanisms in the surroundings of the central supermassive black hole.\\

Markarian 421 (\object{Mkn~421}) is the closest known BL~Lac blazar (at a redshift of $0.031$), and the first discovered in the high
and very-high energy (VHE) ranges. It was first detected as a weak source by the EGRET instrument up to a few $\mathrm{GeV}$ during summer
1991 (\cite{Lin92}). Eight months later, the Whipple Observatory
detected a clear signal from this object between $0.5$ and $1.5\:\mathrm{TeV}$ (\cite{Punch92}).
Since then, \object{Mkn~421} has been confirmed many times as a VHE source by various atmospheric Cherenkov imaging telescopes, e.g. again by the
Whipple Observatory (\cite{Krennrich97}), by the stereoscopic system of HEGRA
(\cite{Petry96}) and by the CAT (Cherenkov Array at Th\'emis) ex\-pe\-ri\-ment (\cite{Piron98,Piron99b}).
Along with Markarian~501 (\object{Mkn~501}), \object{Mkn~421} has thus become one of the two extragalactic sources of the Northern hemisphere which has
been firmly established in the VHE range. It has been also one of the most studied blazars and the target of several multi-wavelength
observation campaigns from the radio band to the $\gamma$-ray range
(see, e.g., \cite{Macomb95}; \cite{Buckley96}; \cite{Takahashi96}; \cite{Takahashi99,Takahashi00}; \cite{Charlot98}; \cite{Maraschi99}).
Recently, a major step was achieved during the 1999--2000 winter by the CELESTE atmospheric Cherenkov sampling experiment, which detected
\object{Mkn~421} for the first time around $50\:\mathrm{GeV}$, fil\-ling the last energy gap still remaining on this source between satellites and ground-based
detectors. These observations, made in part simultaneously with the CAT telescope, have been presented by~\cite{Holder01}.\\

In this paper we concentrate on the temporal variability and the spectral properties of \object{Mkn~421} above $250\:\mathrm{GeV}$, as seen by the CAT experiment since it began operation in autumn 1996.
Sect.~\ref{SecDet} describes the detector and the analysis methods
used to extract the signal and the spectra. The \object{Mkn~421} data sample, the light curves up to June 2000
and the corresponding spectra are presented in Sect.~\ref{SecRes}. We discuss our results in Sect.~\ref{SecDiscuss},
comparing them with those from other ground-based atmospheric Cherenkov telescopes, and with those
obtained on \object{Mkn~501}. The conclusions are given in Sect.~\ref{SecConcl}.

\section{Experimental setup and data analysis}
\label{SecDet}
\subsection{Telescope characteristics}
\label{SecDetSubSecDet}
Located on the site of the former solar plant at Th\'emis in the French Pyr\'en\'ees ($2\degr$ East, $42\degr$ North, altitude $1650\:$m
above sea level), the CAT telescope (\cite{Barrau98}) uses the Cherenkov imaging technique. Its $17.8\:$m$^2$ mirror collects
the Cherenkov light emitted by the secondary particles produced during the development of atmospheric showers and forms their image
in its focal plane. The CAT camera has a full field of view of $4.8\degr$ but the present study is limited to the
$3.1\degr$ fine-grained inner part, which is comprised of $546$ phototubes with $0.12\degr$ angular diameter, arranged in an hexagonal
matrix. This very high definition, combined with fast electronics whose integration time ($12\:$ns) matches the
Cherenkov flash's short duration, is an efficient solution of the two main
problems of night-sky and cosmic-ray backgrounds which confront such $\gamma$-ray
detectors.
Firstly, the detection threshold, which is determined by the night-sky noise, is as low as
$250\:\mathrm{GeV}$ at Zenith\footnote{The threshold is defined here following the usual convention for atmospheric
Cherenkov detectors, i.e. as the energy at which the differential $\gamma$-ray
rate is maximum at the trigger level for a Crab-like spectrum. As shown in Fig.~\ref{FigAcceff}, it is
somewhat higher after the event selection applied at the analysis level.}.
Secondly, the capability of rejecting the huge cosmic-ray background (protons, nuclei) is improved by an
accurate analysis based on the comparison of individual images with theoretical mean $\gamma$-ray images. This method, very briefly
reviewed in Sect.~\ref{SecDetSubSecSig}, is specific to the CAT experiment and has been des\-cri\-bed in detail in~\cite{LeBohec98}.

\subsection{Gamma-ray signal extraction}
\label{SecDetSubSecSig}
In order to improve the hadronic rejection and stabilize the background level near the detection threshold,
and to compensate for possible slight changes in the detector response between different epochs of observation, we
eliminate the noi\-siest pixels and require the fourth-brightest-pixel's charge
in the ima\-ge $Q_\mathrm{4}$$>$$3\:$p.e. (photo-electrons) and the image's total charge
$Q_{\mathrm{tot}}$$>$$30\:$p.e.

An efficient discrimination between $\gamma$ and hadron-induced showers is then achieved by looking at the shape and the
orientation of the images. Since $\gamma$-ray images are rather thin and ellipsoidal
while hadronic images are more irregular, a first cut is applied which selects images with a ``$\gamma$-like''
shape; it is based on a $\chi^2$ fit to a mean light distribution predicted from electromagnetic showers, and a
probability $\mathcal{P}({\chi^2})$$>$$0.35$ is required. In addition, since $\gamma$-ray images are expected to point towards the
source angular position
in the focal plane whereas cosmic-ray directions are isotropic, a second cut $\alpha$$<$$6\degr$ is used in the
case of a point-like source, where the pointing angle $\alpha$ is defined as the angle at the image barycentre
between the actual source angular position and the source position as reconstructed by the
fit. As a result, this procedure rejects $99.5$\% of hadronic events while keeping $40$\% of
$\gamma$-ray events; the \object{Crab nebula}, which is generally considered as the standard
candle for VHE $\gamma$-ray astronomy, is detected at a $4.5\:\sigma$ level in one hour.\\

Fig.~\ref{FigAlplot} shows the $\alpha$ distributions obtained from two data samples taken on \object{Mkn~421}, for ON and OFF-source
observations (the latter being taken at the same telescope elevation in order to monitor the hadronic background),
and the cor\-res\-pon\-ding distributions for $\gamma$-rays obtained by ``ON--OFF'' subtraction (bottom-left insets).
The signal is clearly seen in the direction of the source (small $\alpha$),
though the direction of some $\gamma$-rays is mis-identified, giving a small signal at $\alpha$$\sim$$180\degr$.
As stated above, the $\chi^2$ fit
also allows the angular origin of $\gamma$-ray events to be determined with good accuracy as it uses the information contained in the
images' asymmetrical longitudinal light profile. In Fig.~\ref{FigAlplot}, the bottom-right insets show the si\-gni\-fi\-can\-ce map of
$\gamma$-ray event excesses: the angular resolution {\it per event} is $0.11\degr$ (i.e.,
of the order of the pixel size), allowing a bright source to be localized with
an accuracy better than $1\arcmin$ (dominated by systematics).

\subsection{Spectral analysis}
\label{SecDetSubSecSp}
\begin{figure}[t]
\begin{center}
\epsfig{file=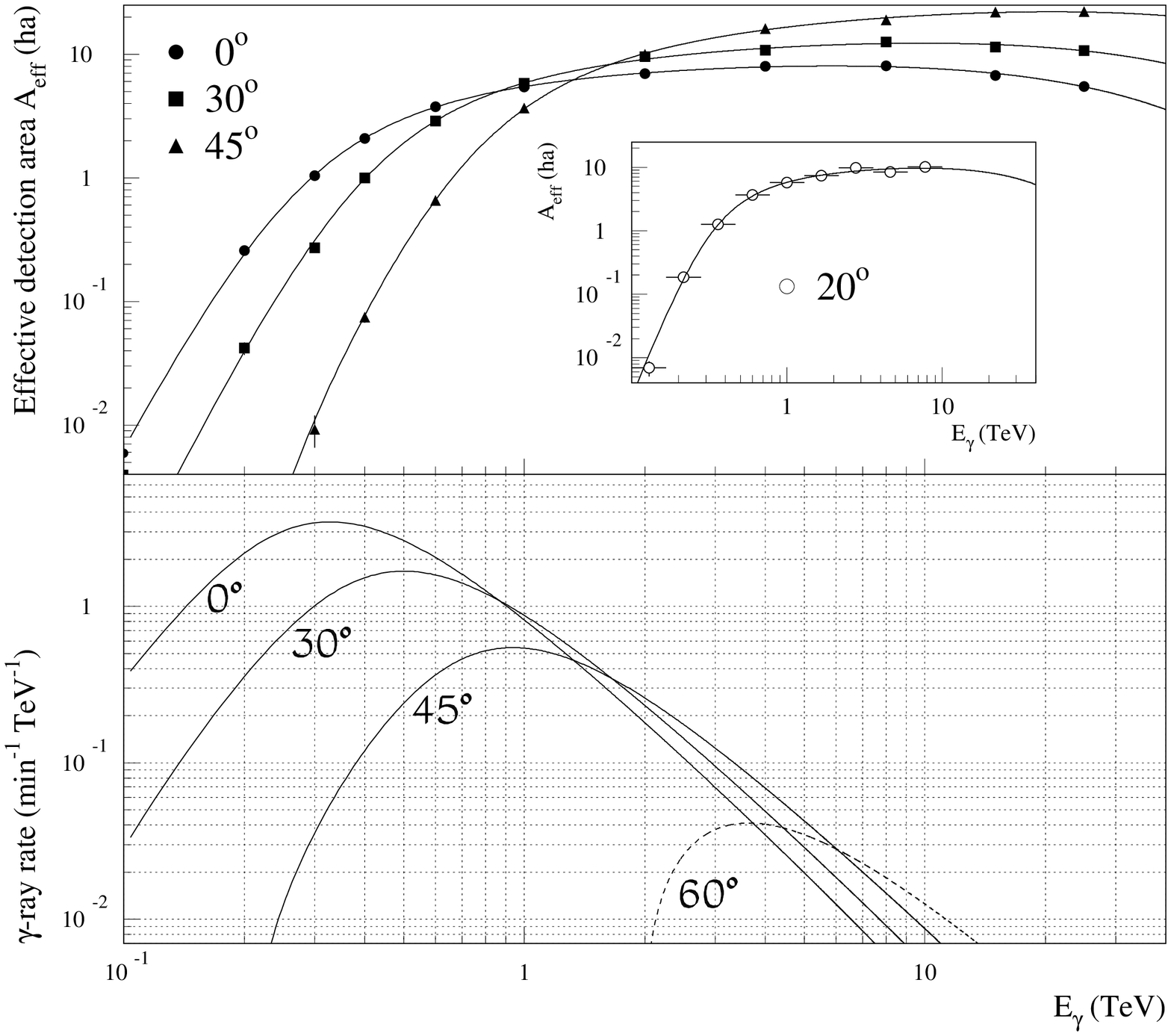,width=\linewidth,clip=
,bbllx=15pt,bblly=30pt,bburx=660pt,bbury=600pt}
\caption[]{CAT effective detection area (top) and differential trigger rate (bottom) for $\gamma$-ray showers
after event selection.\\
\underline{Top:} each point in the main panel is a simulation, while full lines come from an analytical 2D-formula
$\mathcal{A}_\mathrm{eff}(\theta_{\mathrm z},E_\gamma)$ allowing interpolation over zenith angle and energy.
The inset compares the prediction of this interpolation for
$\theta_{\mathrm z}$$=$$20\degr$ with an independent simulation (open circles);\\
\underline{Bottom:} differential $\gamma$-ray trigger rate ($\frac{\mathrm{d}\phi}{\mathrm{d}E}$$\times$$\mathcal{A}_\mathrm{eff}$), for
a typical spectrum $\frac{\mathrm{d}\phi}{\mathrm{d}E}$$=$$3.0\:E_{\mathrm{TeV}}^{-2.55}\times$10$^{-11}\:$cm$^{-2}$\,s$^{-1}$\,TeV$^{-1}$ and different values of $\theta_{\mathrm z}$. The dashed curve at $60\degr$ is only
indicative, as the interpolation $\mathcal{A}_\mathrm{eff}$ must be still validated for large zenith angles ($\theta_{\mathrm z}$$\gtrsim$$45\degr$).}
\label{FigAcceff}
\end{center}
\end{figure}
\begin{figure*}[t]
\begin{center}
\epsfig{file=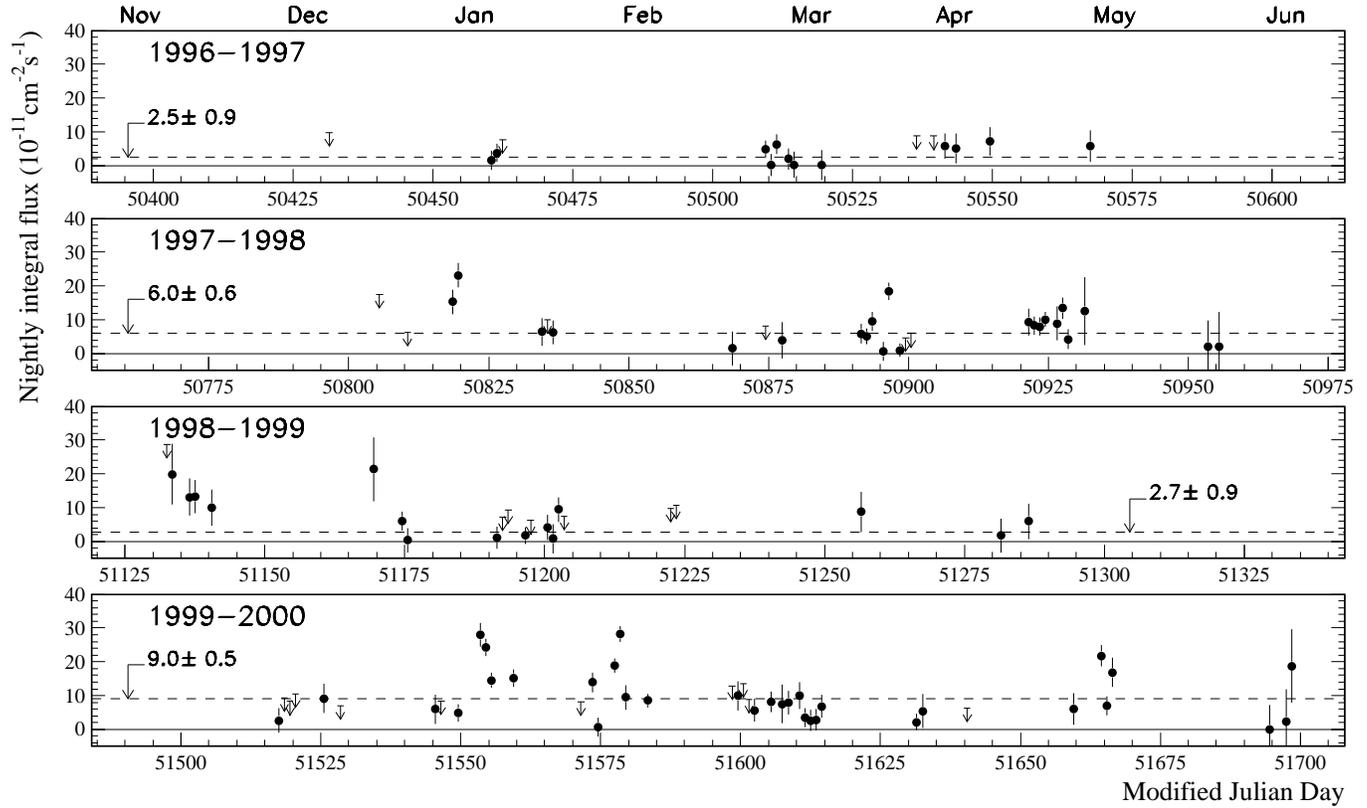,width=\linewidth,clip=
,bbllx=30pt,bblly=35pt,bburx=460pt,bbury=295pt}
\caption[]
{\object{Mkn~421} nightly-averaged integral flux above $250\:\mathrm{GeV}$ between December, 1996, and June, 2000.
The $\gamma$-ray effective area has been weighted using a differential index of $-2.9$,
in order to estimate the integral flux for observations far from the Zenith (see appendix~\ref{AnnSubSecFlux}).
Arrows stand for $2\:\sigma$ upper-limits when no signal was recorded, and dashed lines show the mean flux for each
observation year.
}
\label{Fig421CL}
\end{center}
\end{figure*}
VHE $\gamma$-ray spectra result from particle acceleration processes and thus
they are expected to steepen above a given energy; this
combines with the energy resolution currently achieved by imaging Cherenkov
atmospheric detectors ($20$\% at best) to cause a considerable event flow into higher {\it estimated} energy intervals.
Starting with an observed differential $\gamma$-ray trigger rate, one therefore needs a global forward-folding method,
using the knowledge of the detector response ($\gamma$-ray effective detection area, energy resolution), {\it as well as a parameterization
of the spectral shape}.
Therefore, we have chosen a maximum likelihood method which directly provides relevant physical results for
the present problem, namely the values of the most probable spectral parameters and their covariance matrix.

The image analysis described in Sect.~\ref{SecDetSubSecSig} also yields the energy of each hypothesised $\gamma$-ray shower.
The spectral analysis presented below involves the exact energy-resolution function $\Upsilon$, which is characterised by a r.m.s. of
$22$\% (independent of energy) and which includes possible bias in energy reconstruction close to the detection threshold.
This function has been determined by detailed Monte-Carlo simulations of the telescope response, as
has the effective detection area $\mathcal{A}_\mathrm{eff}$, which includes the effect of event-selection efficiency (see Fig.~\ref{FigAcceff}).
The simulations have been checked and calibrated on the basis of several observables, especially
by using muon rings and the nearly-pure $\gamma$-ray signal from the highest flare of \object{Mkn~501} in April 1997
(\cite{Piron99a}).\\

With typical statistics of $\sim$$1000$ $\gamma$-ray events and signal-to-background ratio of $\sim$$0.4$
(as obtained on the \object{Crab ne\-bu\-la}), a spectrum can be determined with reasonable accuracy as follows.
First we define a set $\displaystyle\{\Delta_{i_z}\}$$\equiv$$\{[\theta^{\mathrm{min}}_{i_z}, \theta^{\mathrm{max}}_{i_z}]\}_{i_z=1,n_z}$
of $n_z$ zenith angle bins, with a width (between $0.02$ and $0.04$ in cosine) small enough compared to the variation scale of $\Upsilon$ and
$\mathcal{A}_\mathrm{eff}$; $\Delta_1$ corresponds to the transit of the source at Th\'emis, and $\Delta_{n_z}$ to the maximum angle fixed by the data sample.
Then we define $n_e$ estimated energy bins
$\displaystyle\{\Delta_{i_e}\}$$\equiv$$\{[\widetilde{E}^{\mathrm{min}}_{i_e}, \widetilde{E}^{\mathrm{max}}_{i_e}]\}_{i_e=1,n_e}$,
with a width ($\geq$$0.2$ in $\log_{10}E_{\mathrm{TeV}}$) at least twice as large as the typical width of the function $\Upsilon$.
The maximum energy $\widetilde{E}^{\mathrm{max}}_{n_e}$ is fixed by the avai\-la\-ble statistics.
Finally, we define a set of bins
$\displaystyle\{\Delta_{i_z, i_e}\}$$\equiv$$\{\Delta_{i_z}\otimes\Delta_{i_e}\}_{i_z=1,n_z;i_e=i_{e1}(i_z),n_e}$;
for each $\Delta_{i_z}$ bin, the lowest energy (and thus the bin $\Delta_{i_{e1}(i_z)}$) is determined by the telescope detection threshold
which increases with zenith angle (see Fig.~\ref{FigAcceff}).

Within each $\Delta_{i_z, i_e}$ 2D-bin, the number of events pas\-sing the selection cuts is determined separately for all ON and
OFF-source data, and the maximum-likelihood estimation of the spectral parameters is performed following the procedure
detailed in appendix~\ref{AnnSubSecSp}. The likelihood-function expression does not rely on a straightforward
``ON--OFF'' subtraction as in usual spectral analyses, but on the {\it
respective} Poissonian distributions of ON and OFF events. In
particular, this allows possible low statistics to be treated in a rigorous manner. No hypothesis is required on the background (OFF) shape,
but two hypotheses are successively considered for the differential $\gamma$-ray spectrum $\displaystyle\frac{\mathrm{d}\phi}{\mathrm{d}E}$: {\it i)}
a simple power law, $\phi_0^\mathrm{pl}\:E_{\mathrm{TeV}}^{-\gamma^\mathrm{pl}}$ (hyp. $\mathcal{H}^\mathrm{pl}$), which is often a good approximation, at least within
a restricted energy range (over one or two orders of magnitude),
and {\it ii)} a curved shape, $\phi_0^\mathrm{cs}\:E_{\mathrm{TeV}}^{-(\gamma^\mathrm{cs}+\beta^\mathrm{cs}\log_{10}E_{\mathrm{TeV}})}$ (hyp. $\mathcal{H}^\mathrm{cs}$).
The latter parameterization, previously used by the Whipple group for the study of \object{Mkn~421} and \object{Mkn~501} (\cite{Krennrich99a}),
corresponds to a parabolic law in a $\log(\nu F(\nu))$ {\it vs.} $\log (\nu)$ re\-pre\-sen\-ta\-tion, where $\displaystyle\nu F(\nu)\equiv E^2\frac{\mathrm{d}\phi}{\mathrm{d}E}$
and $E=h\nu$.

The relevance of $\mathcal{H}^\mathrm{pl}$ with respect to $\mathcal{H}^\mathrm{cs}$ is estimated from the likelihood ratio of the two hypotheses, which is defined as
$\displaystyle\lambda=2\log\left(\frac{\mathcal{L}^\mathrm{cs}}{\mathcal{L}^\mathrm{pl}}\right)$: it behaves (asymptotically) like a $\chi^2$ with one degree of freedom and permits
the search for possible spectral curvature. For each data sample, the spectral law finally retained is given by the most relevant
parameterization of the differential spectrum. In the following, we chose to represent each spectrum as a function of the {\it true}
photon energy by an area corresponding to the $68$\% confidence level contour given by the likelihood method.

\section{Results}
\label{SecRes}
\subsection{Data sample and light curves}
\label{SecResSubSecSample}
The complete data sample consists of observations taken between December, 1996, and June, 2000.
During these periods, the source was systematically observed in a range of zenith angle extending from close to the Zenith up to $45\degr$.
The intensity of the source did not influence the observation strategy.
However, a selection based on criteria requiring clear moonless nights and stable detector operation has been applied:
this leaves a total of $139$~hours of on-source (ON) data, together with $57$~hours on control (OFF) regions.
The different light curves of the four observation pe\-riods are shown in Fig.~\ref{Fig421CL}. We used a
differential index of $-2.9$, which is representative of all spectral measurements presented in Sect.~\ref{SecResSubSecSpectra},
to estimate the integral flux above $250\:\mathrm{GeV}$ for all data, especially those taken far from the Zenith: this procedure is
detailed in appendix~\ref{AnnSubSecFlux}.

As can be seen in Fig.~\ref{Fig421CL}, the flux of \object{Mkn~421} changed significantly between
1996--97 and 1997--98: almost quiet during the first period (with a mean flux $\Phi_{>250\:\mathrm{GeV}}$$=$$2.5\pm0.9\times10^{-11}\:\mathrm{cm}^{-2}\:\mathrm{s}^{-1}$),
the source showed a higher mean activity during the second period ($\Phi_{>250\:\mathrm{GeV}}$$=$$6.0\pm0.6\times10^{-11}\:\mathrm{cm}^{-2}\:\mathrm{s}^{-1}$), with small bursts
in January and March sometimes showing up in excess of the steady flux from the \object{Crab nebula} (which is
$\Phi^\mathrm{CN}$$\equiv$$14.10\pm0.35\times10^{-11}\:\mathrm{cm}^{-2}\:\mathrm{s}^{-1}$ above $250\:\mathrm{GeV}$, see~\cite{PironThese}).
In 1998--99, the mean VHE emission of \object{Mkn~421} ($\Phi_{>250\:\mathrm{GeV}}$$=$$2.7\pm0.9\times10^{-11}\:\mathrm{cm}^{-2}\:\mathrm{s}^{-1}$)
decreased to a level comparable to that of 1996--97. In spite of some activity detected during the winter, the weather conditions
in Th\'emis caused a very sparse source coverage. Nevertheless in the beginning of 2000
\object{Mkn~421} showed a remarkable increase in activity, exhibiting a series of huge
bursts. As seen in Fig.~\ref{Fig421CL}, the bursts recorded in January and February
2000 clearly appear as the highest ever seen by CAT from this source
in four years with a nightly-averaged integral flux culminating at $\sim$$2\:\Phi^\mathrm{CN}$ and a large night-to-night
variability.

\begin{figure}[t]
\begin{center}
\vbox{
\hbox{
\epsfig{file=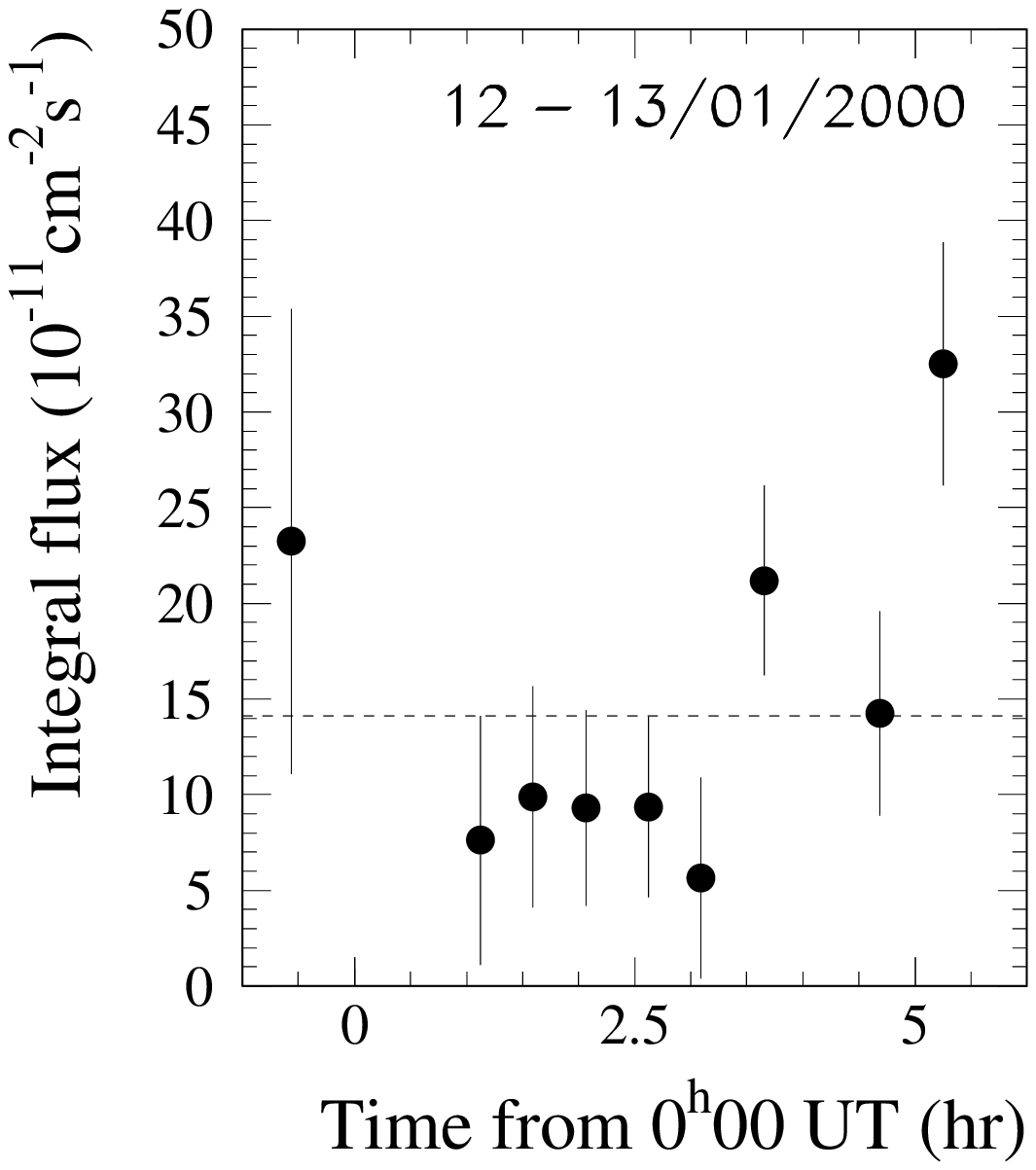,width=.97\linewidth,clip=
,bbllx=10pt,bblly=25pt,bburx=800pt,bbury=385pt}
\hspace*{-0.57\linewidth}
\parbox[h]{0.55\linewidth}{
\vspace*{-4.5cm}
\caption[]{\object{Mkn~421} integral flux above $250\:\mathrm{GeV}$ during the night of January 12--13, 2000 (MJD 51555/56, left panel), and
between 3 and 6 February, 2000 (MJD 51577 to 51580, lower panels).
Each point stands for a $\sim$$30\:\mathrm{min}$ observation and the dashed lines show the \object{Crab nebula} level emission $\:\Phi^\mathrm{CN}$.
}
\label{FigVar}
}
}
\epsfig{file=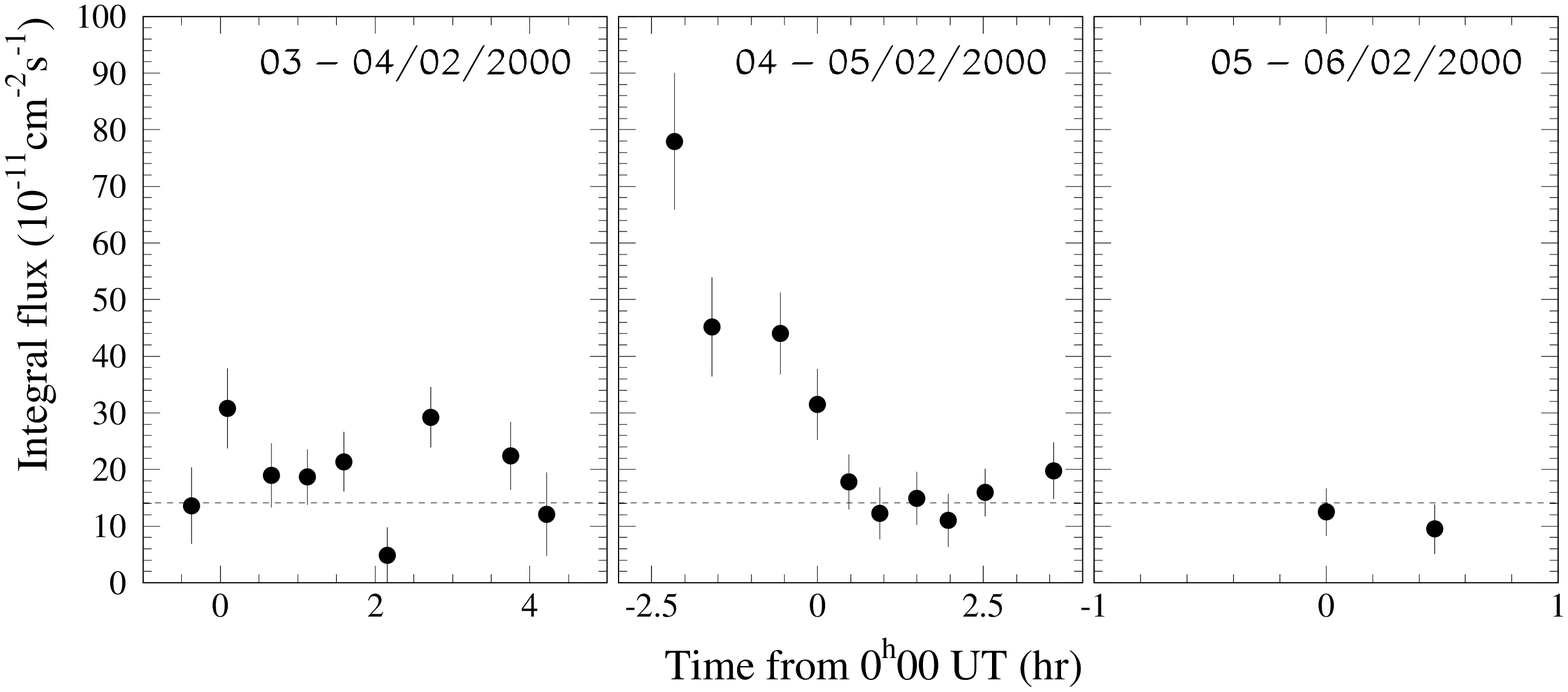,width=.97\linewidth,clip=
,bbllx=10pt,bblly=25pt,bburx=800pt,bbury=385pt}
}
\end{center}
\end{figure}

VHE intra-night variability was also observed on a few occasions. For instance during the night of January 12--13,
the source intensity increased by a factor of $3.8$ in $\sim$$2$~hours, from $\sim$$0.6\:\Phi^\mathrm{CN}$ to
$\sim$$2.3\:\Phi^\mathrm{CN}$
(with a $\chi^2$ per d.o.f of $2.5$ for the absence of any variation), as can be seen in Fig.~\ref{FigVar} (upper-left panel).
At the bottom of this figure, \object{Mkn~421} light curves are also shown for three nights from the
3$^\mathrm{rd}$ to the 5$^\mathrm{th}$ February. While the fluxes recorded by CAT
during the first and last nights were stable, respectively $\Phi_{>250\:\mathrm{GeV}}$$\simeq$$1.3\:\Phi^\mathrm{CN}$ (over $4$~hours) and
$\Phi_{>250\:\mathrm{GeV}}$$\simeq$$0.7\:\Phi^\mathrm{CN}$ (over $1$~hour), the source activity
changed dramatically in a few hours during the se\-cond night (February 4--5). The CAT telescope started observation while the
source emission was at a level of $5.5\:\Phi^\mathrm{CN}$. This flux is comparable to the historically highest $\mathrm{TeV}$ flux ever
recorded, i.e., that of \object{Mkn~501} during the night of April 16$^\mathrm{th}$, 1997
(\cite{Djannati99}). In spite of the low source elevation ($\theta_{\mathrm z}$$=$$44\degr$) $124$ $\gamma$-ray
events with a signal significance of $6.5\:\sigma$ were detected during the first $30$ minutes of observation.
This may be compared to the $838$ $\gamma$-ray events and significance of $13.8\:\sigma$
obtained during the whole night (see Fig.~\ref{FigAlplot}). After this first episode, the source intensity was reduced by a
factor of $2$ in $1$~hour and by a factor of $5.5$ in $3$~hours. In Fig.~\ref{FigVar}, each point stands for a
$\sim$$30\:\mathrm{min}$ observation but a finer binning in time does not show any additional interesting features, confirming that CAT
started observation after the flare maximum.

\subsection{1998 and 2000 time-averaged spectra}
\label{SecResSubSecSpectra}
\begin{figure*}[t!]
\begin{center}
\hbox{
\textbf{(a)}
\hspace*{-.03\linewidth}
\epsfig{file=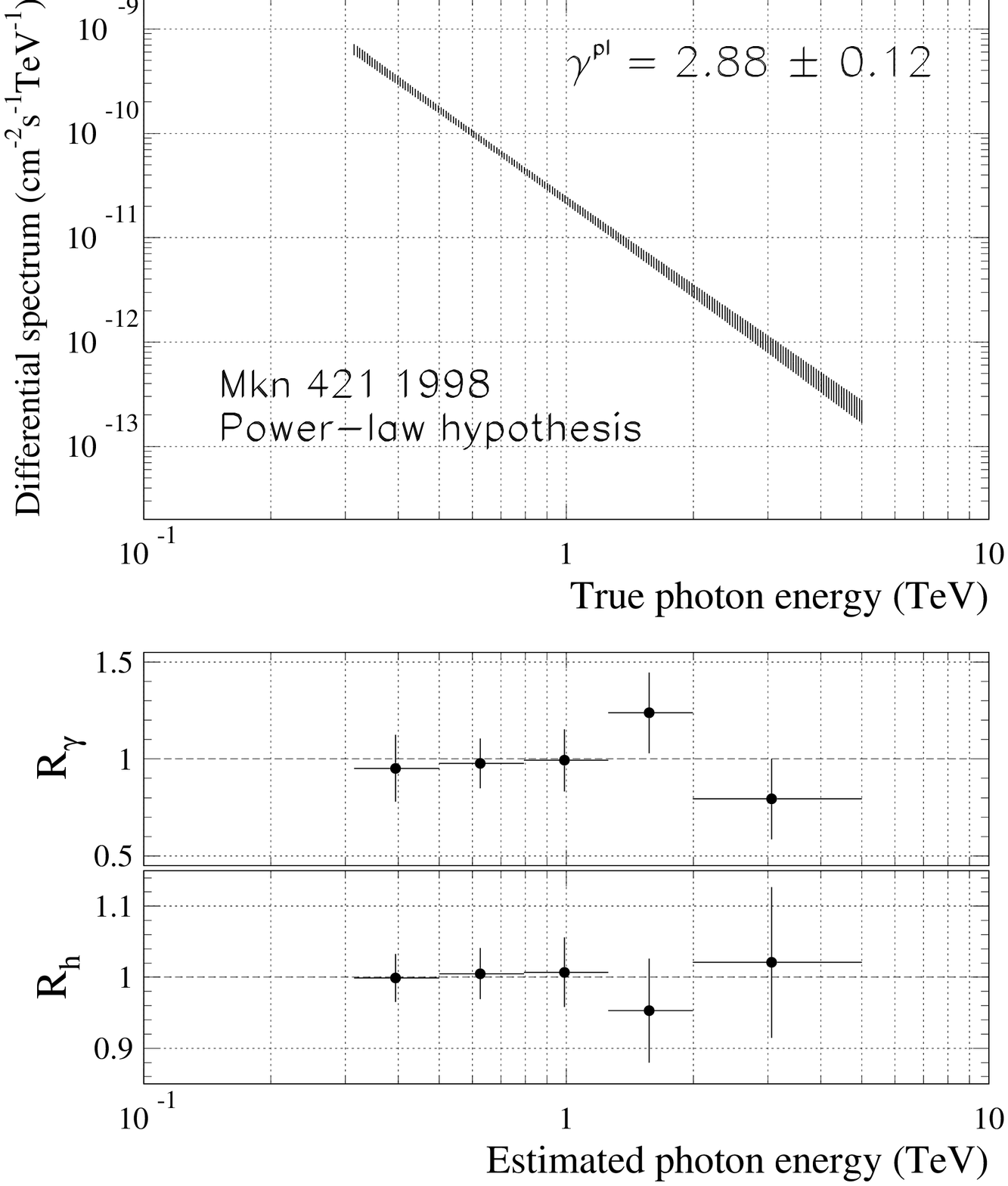,width=.47\linewidth,clip=
,bbllx=5pt,bblly=145pt,bburx=565pt,bbury=800pt}
\hspace*{.04\linewidth}
\textbf{(b)}
\hspace*{-.03\linewidth}
\epsfig{file=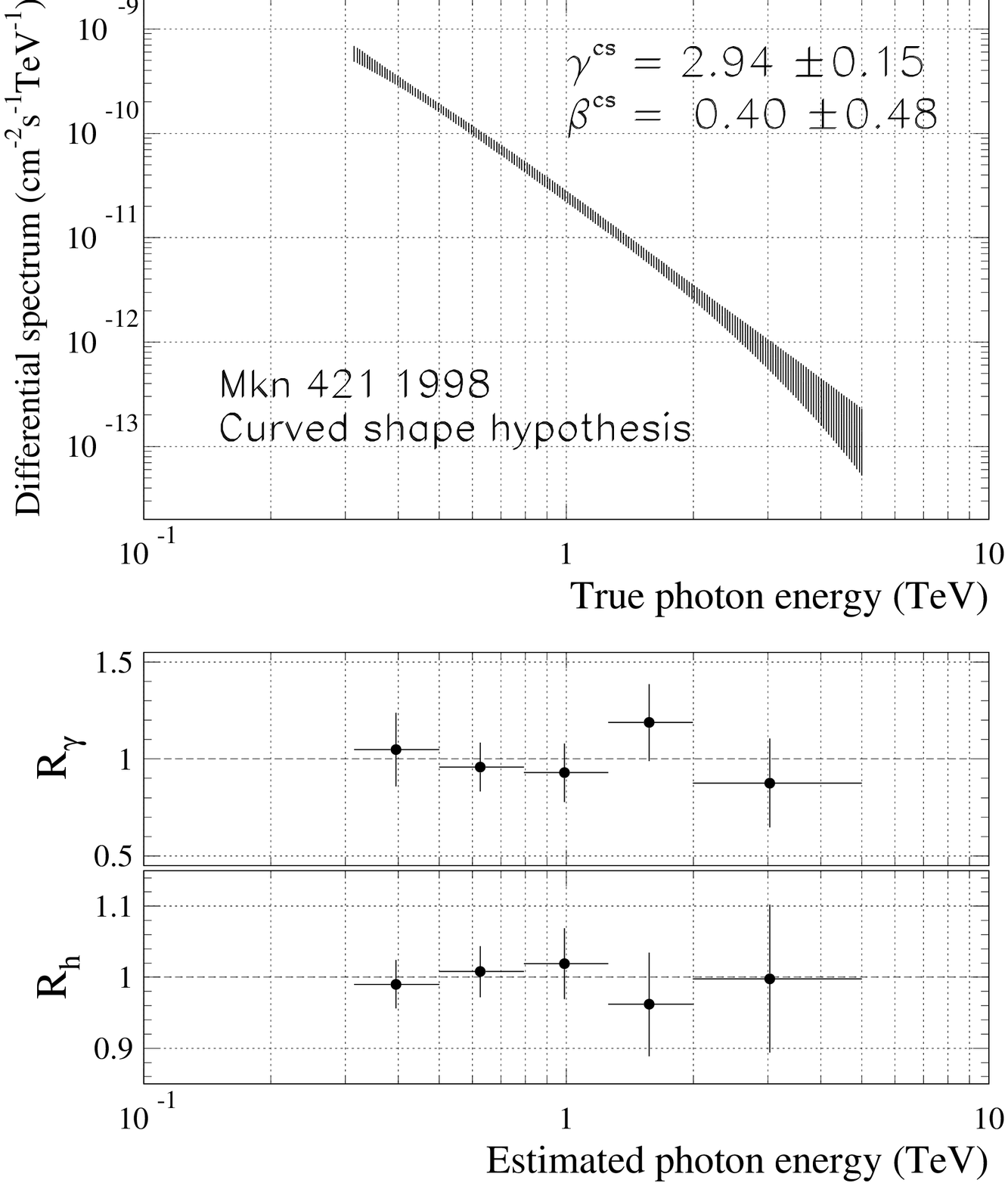,width=.47\linewidth,clip=
,bbllx=5pt,bblly=145pt,bburx=565pt,bbury=800pt}
}
\vspace*{1.5cm}
\hbox{
\textbf{(c)}
\hspace*{-.03\linewidth}
\epsfig{file=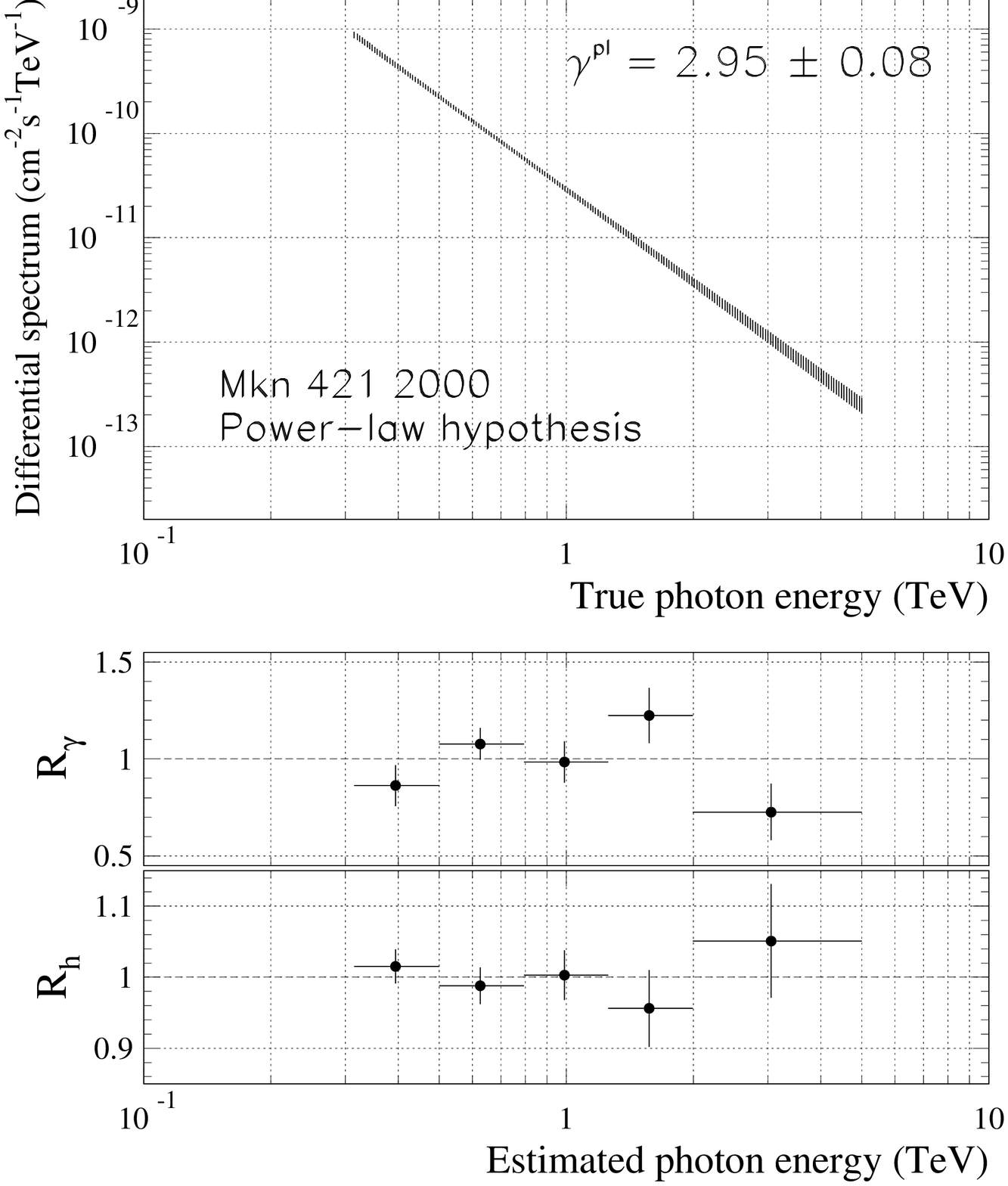,width=.47\linewidth,clip=
,bbllx=5pt,bblly=145pt,bburx=565pt,bbury=800pt}
\hspace*{.04\linewidth}
\textbf{(d)}
\hspace*{-.03\linewidth}
\epsfig{file=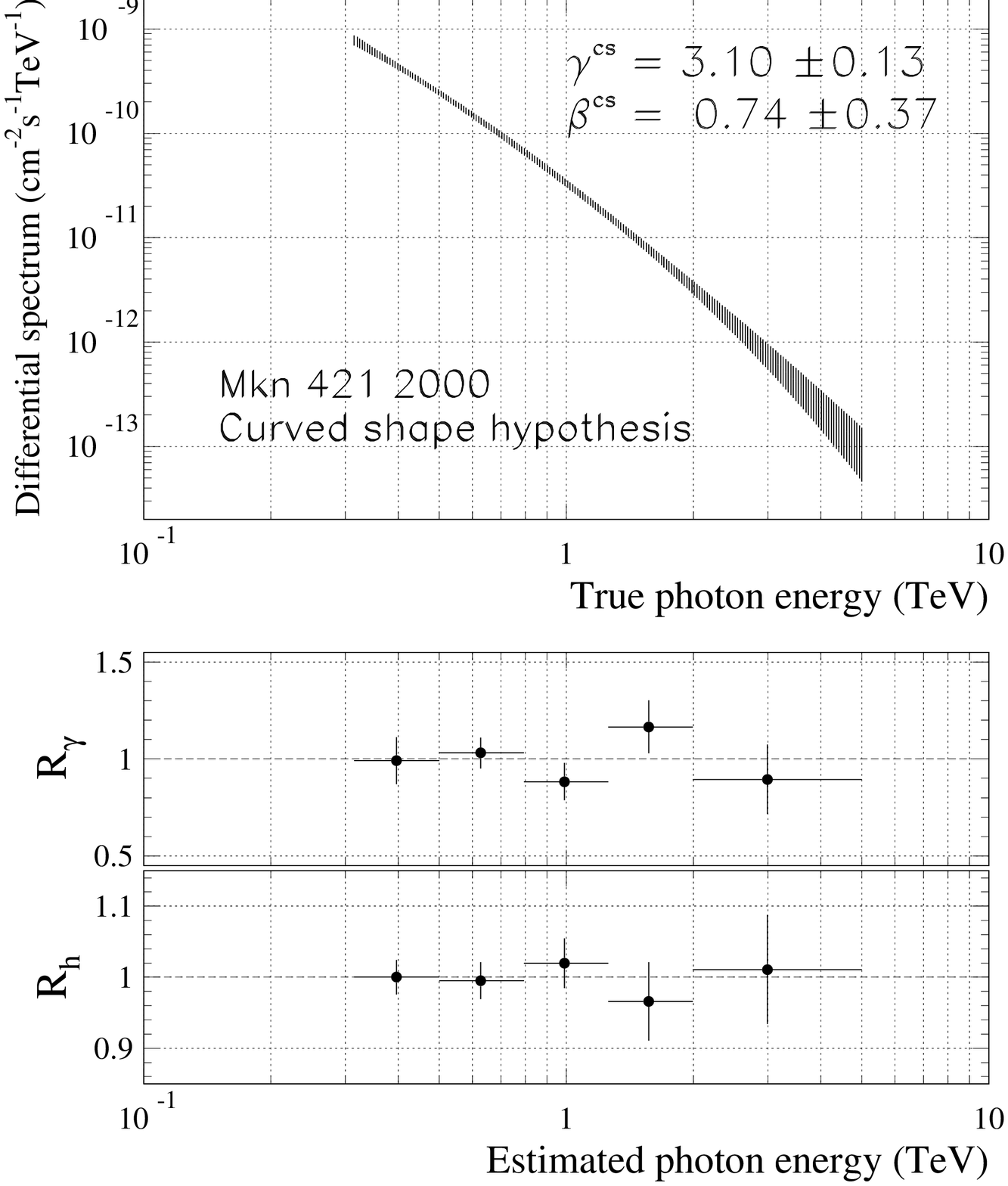,width=.47\linewidth,clip=
,bbllx=5pt,bblly=145pt,bburx=565pt,bbury=800pt}
}
\caption[]
{
\object{Mkn~421} time-averaged spectra between $0.3$ and $5.0\:\mathrm{TeV}$ in 1998 and 2000, for the
power-law and curved shape hypotheses.
The areas show the $68$\% confidence level contour given by the likelihood method.
In each of the four panels, the two lower plots give the ratio, in each bin of {\it estimated} energy, of the predicted number of events
to that which is observed both for the $\gamma$-ray signal ($R_\gamma$) as well as for the hadronic background ($R_\mathrm{h}$).
}
\label{Fig421sp}
\end{center}
\end{figure*}

\newcommand{\zzs}[1]{\scriptstyle{#1}}
\newcommand{\zza}[4]{$\zzs{#1}-\zzs{#2}$&$\zzs{#3}\pm\zzs{#4}$}
\newcommand{\zzb}[5]{$\zzs{#1}\pm\zzs{#2}$&$\zzs{#3}\pm\zzs{#4}$&$\zzs{#5}$}
\newcommand{\zzc}[9]{$\zzs{#1}\pm\zzs{#2}$&$\zzs{#3}\pm\zzs{#4}$&$\zzs{#5}\pm\zzs{#6}$&$\zzs{#7}$&$\zzs{#8}\pm\zzs{#9}$}

\begin{table*}[t!]
\begin{center}
\caption[]
{
Characteristics of the \object{Mkn~421} spectra obtained in this paper. For each spectrum we indicate the observation period,
the total energy band used in the likelihood method, $\Delta\widetilde{E_\gamma}$, the total observed number of $\gamma$-ray events,
$S_\gamma$, the spectral parameters obtained in the $\mathcal{H}^\mathrm{pl}$ and $\mathcal{H}^\mathrm{cs}$ hypotheses, and the likelihood ratio $\lambda$.
We also quote the decorrelation energy in the $\mathcal{H}^\mathrm{pl}$ hypothesis, $E_\mathrm{d}$, the energy $E^\mathrm{cs}_0$ at which the
energy-dependent exponent
$\gamma^\mathrm{cs}_l(E_{\mathrm{TeV}})$$=$$\gamma^\mathrm{cs}$$+$$\beta^\mathrm{cs}\log_{10}E_{\mathrm{TeV}}$
has a minimal error in the $\mathcal{H}^\mathrm{cs}$ hypothesis (see appendix~\ref{AnnSecSp421SubSecCov}),
and the corresponding value $\gamma^\mathrm{cs}_0\equiv\gamma^\mathrm{cs}_l(E^\mathrm{cs}_0)$.
The energies are given in $\mathrm{TeV}$, and the flux constants in units of 10$^{-11}\:$cm$^{-2}$\,s$^{-1}$\,TeV$^{-1}$.
}
\begin{tabular}{cccccccccccc}
\hline
\footnotesize{Period}&
$\zzs{\Delta\widetilde{E_\gamma}\;(\mathrm{TeV})}$&$\zzs{S_\gamma}$&
$\zzs{\phi_0^\mathrm{pl}}$&$\zzs{\gamma^\mathrm{pl}}$&$\zzs{E_\mathrm{d}}$&
$\zzs{\phi_0^\mathrm{cs}}$&$\zzs{\gamma^\mathrm{cs}}$&$\zzs{\beta^\mathrm{cs}}$&$\zzs{E^\mathrm{cs}_0}$&$\zzs{\gamma^\mathrm{cs}_0}$&
$\zzs{\lambda}$\\
\hline
$\zzs{1998}$&
\zza{0.3}{5.0}{735}{57}&
\zzb{2.29}{0.20}{2.88}{0.12}{0.69}&
\zzc{2.51}{0.32}{2.94}{0.15}{0.40}{0.48}{0.73}{2.88}{0.14}&
$\zzs{0.75}$\\
$\zzs{2000}$&
\zza{0.3}{5.0}{1424}{71}&
\zzb{2.90}{0.18}{2.95}{0.08}{0.63}&
\zzc{3.34}{0.28}{3.10}{0.13}{0.74}{0.37}{0.61}{2.94}{0.10}&
$\zzs{4.90}$\\
$\zzs{4-5/02/2000}$&
\zza{0.3}{2.0}{609}{40}&
\zzb{3.16}{0.34}{2.82}{0.15}{0.57}&
\zzc{3.25}{0.37}{3.08}{0.39}{0.63}{0.81}{0.37}{2.81}{0.17}&
$\zzs{0.61}$\\
\hline
\end{tabular}
\label{TabSpRes}
\end{center}
\end{table*}
The data used in this section consist of a series of
$\sim$$30\:\mathrm{min}$ acquisitions for which a $\gamma$-ray signal with significance greater
than $3\sigma$ was recorded, and they have been further limited to zenith angles $\theta_{\mathrm z}$$<$$28\degr$, i.e., to a configuration for
which the detector calibration has been fully completed. The spectral study is thus based on $6.2$~hours of on-source (ON) data taken in
1998 and $8.4$~hours in 2000. Though this data selection reduces somewhat the total number of $\gamma$-ray events, it provides a high
signal-to-noise ratio, minimizes systematic effects, and allows a robust spectral determination. Concerning
systematic effects, another favourable factor is the low night-sky background in the field of view due to the lack of
bright stars around the source.

Systematic errors are thus mainly due to the uncertainty on the absolute energy scale, which comes
from possible variations of the atmosphere transparency and light-collection efficiencies during the observation periods.
To a lesser extent, they are also due to limited Monte-Carlo statistics in the determination of the effective detection area.
These errors, assumed to be the same for all spectra, are implicitly considered in the following and they have been estimated from
detailed simulations (\cite{PironThese}):
$(\Delta\phi_0/\phi_0)^\mathrm{syst}$$=$$\pm$$20$\%, $(\Delta\gamma)^\mathrm{syst}$$=$$\pm$$0.06$, and
$(\Delta\beta)^\mathrm{syst}$$=$$\pm$$0.03$.\\

The 1998 and 2000 time-averaged spectra are shown in Fig.~\ref{Fig421sp}, both in the
power-law and curved shape hypotheses. The statistics used for their extraction are detailed in
appendix~\ref{AnnSecSp421SubSecStat}, the spectral parameters are summarized in Table~\ref{TabSpRes}, and their
covariance matrices are given in appendix~\ref{AnnSecSp421SubSecCov}. In each panel of Fig.~\ref{Fig421sp},
the two lower plots give the ratio, in each bin of {\it estimated} energy, of the predicted number
of events to that which is observed both for the $\gamma$-ray signal ($R_\gamma$) as well as for the hadronic background
($R_\mathrm{h}$).
This is another means to check the validity of the parameters estimation, and to compare between the two hypotheses
on the spectral shape.

As can be seen in Fig.~\ref{Fig421sp}a, the power law accounts very well for the 1998 time-averaged spectrum. The likelihood
ratio value is low ($\lambda$$=$$0.75$, corresponding to a chance pro\-ba\-bi\-li\-ty of $0.39$), and the curvature term is compatible with zero
($\beta^\mathrm{cs}$$=$$0.40$$\pm$$0.48^\mathrm{stat}$). Thus, we find the following differential spectrum:\\
\begin{displaymath}
\frac{\mathrm{d}\phi}{\mathrm{d}E}=(2.29\pm0.20^\mathrm{stat}\pm0.46^\mathrm{syst})10^{-11}\:\mathrm{cm}^{-2}\:\mathrm{s}^{-1}\:\mathrm{TeV}^{-1}
\end{displaymath}
\begin{displaymath}
\times E_{\mathrm{TeV}}^{-2.88\pm0.12^\mathrm{stat}\pm0.06^\mathrm{syst}}.
\end{displaymath}

On the contrary, Fig.~\ref{Fig421sp}d shows some evidence for a curvature in the 2000 time-averaged spectrum, with a
relatively high likelihood ratio value ($\lambda$$=$$4.90$, corresponding to a chance probability of $0.027$).
The lower plots of Fig.~\ref{Fig421sp}d, when compared
to those of Fig.~\ref{Fig421sp}c, directly confirm that a curved spectrum provides a better fit to the data.
This curvature is equivalent to an energy-dependent exponent $\gamma^\mathrm{cs}_l(E_{\mathrm{TeV}})$$=$$\gamma^\mathrm{cs}$$+$$\beta^\mathrm{cs}\log_{10}E_{\mathrm{TeV}}$.
Table~\ref{TabSpRes} shows the energy $E^\mathrm{cs}_0$ at which the error on $\gamma^\mathrm{cs}_l(E_{\mathrm{TeV}})$ is minimal, as well as the corresponding
value $\gamma^\mathrm{cs}_0\equiv\gamma^\mathrm{cs}_l(E^\mathrm{cs}_0)$; the latter value is in all cases very close to that of $\gamma^\mathrm{pl}$ (in the absence of curvature),
showing the consistency of the spectral analysis. Thus, we finally retain the following parameterization for the 2000 period:
\begin{displaymath}
\frac{\mathrm{d}\phi}{\mathrm{d}E}=(3.34\pm0.28^\mathrm{stat}\pm0.67^\mathrm{syst})10^{-11}\:\mathrm{cm}^{-2}\:\mathrm{s}^{-1}\:\mathrm{TeV}^{-1}
\end{displaymath}
\begin{displaymath}
\times E_{\mathrm{TeV}}^{-3.10\pm0.13^\mathrm{stat}\pm0.06^\mathrm{syst}-(0.74\pm0.37^\mathrm{stat}\pm0.03^\mathrm{syst})\log_{10}E_{\mathrm{TeV}}}.
\end{displaymath}

\section{Discussion}
\label{SecDiscuss}
\subsection{Comparison with other VHE detections}
\label{SecDiscussSubSecCompExp}
\begin{figure}[t!]
\begin{center}
\epsfig{file=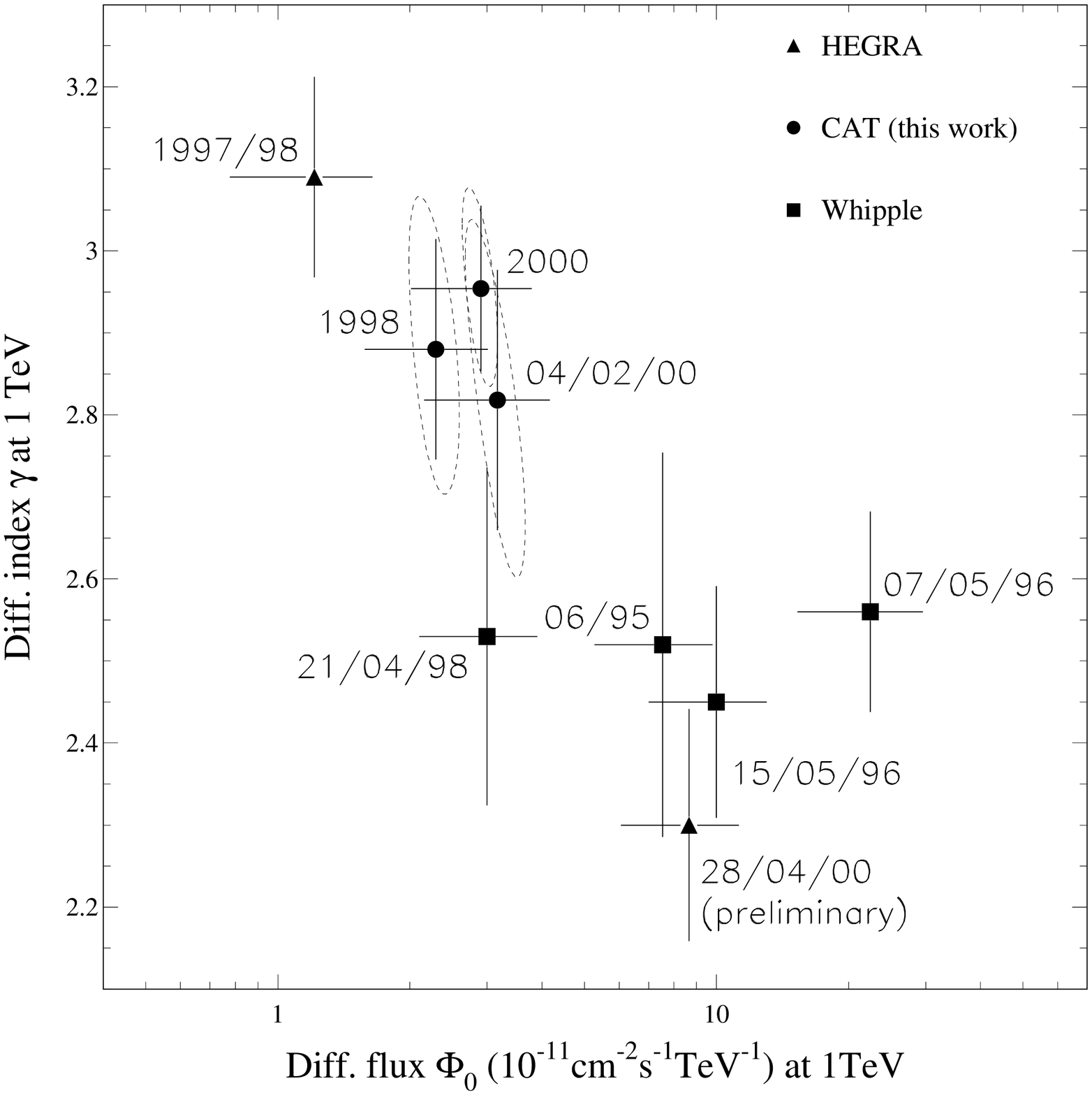,width=\linewidth,clip=
,bbllx=15pt,bblly=15pt,bburx=660pt,bbury=660pt}
\caption[]
{
Compilation of spectral measurements on \object{Mkn~421}, obtained by the HEGRA (\cite{Aharonian99}; \cite{Horns01}),
CAT (this work) and Whipple (\cite{Zweerink}; \cite{Krennrich99a,Krennrich99b}) experiments between 1995 and 2000.
The results are given in the $\{\phi_0^\mathrm{pl},\gamma^\mathrm{pl}\}$ plane of spectral parameters in the power-law hypothesis.
While the dashed lines around CAT points show only the $68$\% confidence level contour (statistical error), the error bars
take account, for {\it all} measurements, of both statistical and systematic errors, the latter being conservative if unknown:
$(\Delta\phi_0^\mathrm{pl}/\phi_0^\mathrm{pl})^\mathrm{syst}$$=$$\pm$$30$\% and $(\Delta\gamma^\mathrm{pl})^\mathrm{syst}$$=$$\pm$$0.10$ were used where these values are
not quoted directly.
The $\chi^2$ per d.o.f corresponding to the absence of any dependence of the spectral index with flux is $3.5$.
}
\label{Fig421cont}
\end{center}
\end{figure}
The energy spectrum of \object{Mkn~421} has been measured in the same energy range by the Whipple Observatory
(\cite{Zweerink}; \cite{Krennrich99a,Krennrich99b}) and by the HEGRA experiment (\cite{Aharonian99}; \cite{Horns01}).
All these experiments have fitted a power law to the spectrum and given the values of $\phi_0^\mathrm{pl}$ and of $\gamma^\mathrm{pl}$ in several
observation periods corresponding to different levels of ac\-ti\-vi\-ty of the source.
Our results for the periods 1998 and 2000 are plotted in Fig.~\ref{Fig421cont} together with their results.
This figure includes both statistical {\it and} systematic errors for {\it all} measurements.
Although CAT data tend to indicate some spectral curvature in 2000, we used the values of $\gamma^\mathrm{pl}$ for this
comparison, given that this parameter is in any case clearly indicative of the energy dependence of the spectrum
($\gamma^\mathrm{cs}_0\simeq\gamma^\mathrm{pl}$, see end of Sect.~\ref{SecResSubSecSpectra}).
In this figure we also added the result obtained during the single night between 4 and 5 February, 2000,
for which the high source intensity (see Sect.~\ref{SecResSubSecSample}) allowed us to extract a spectrum. Since the corresponding
errors are larger (see Table~\ref{TabSpRes}), this single result is fully compatible with the other two.\\

Some caution is necessary when comparing results from  experiments with different systematic errors.
As an example, in the case of a curved spectrum, the value of the ener\-gy $E^\mathrm{cs}_0$
(see Table~\ref{TabSpRes}) corresponding to the minimal error on the spectral exponent $\gamma^\mathrm{cs}_l$
is not necessarily the same for experiments with dif\-fe\-rent energy thresholds.
Possible sys\-te\-ma\-tic effects could be responsible for the dispersion of results shown in Fig.~\ref{Fig421cont}, since,
if one excepts a recent preliminary result from HEGRA (\cite{Horns01}), HEGRA and CAT data favour higher values of
$\gamma^\mathrm{pl}$, whereas data from the Whipple Observatory yield a rather low spectral index.
However, ex\-pe\-ri\-ments showing the largest difference in spectral indices in the figure (Whipple Observatory and HEGRA)
are in good agreement on the spectrum of the Crab nebula (\cite{Hillas98}; \cite{Aharonian00}).
Furthermore, some evidence for a spectral variability has been already suggested on the basis of the Whipple 1995--96
low-flux data (\cite{Zweerink}), with a spectral index $\gamma^\mathrm{pl}$$=$$2.96$$\pm$$0.22$, but
no flux value is quoted in the latter re\-fe\-ren\-ce, precluding any comparison with the results discussed here.

Data from all experiments with the source in both low and high states would be desirable in order to exclude 
that experimental effects are responsible for the dispersion seen in Fig.~\ref{Fig421cont}.
Besides, as different flares may be intrinsically different from each other, the spectral index-flux correlation may not
provide us with complete information, thus stressing again the need for more numerous observations.

\subsection{Comparison with results on \object{Mkn~501}}
\begin{figure}[t]
\begin{center}
\epsfig{file=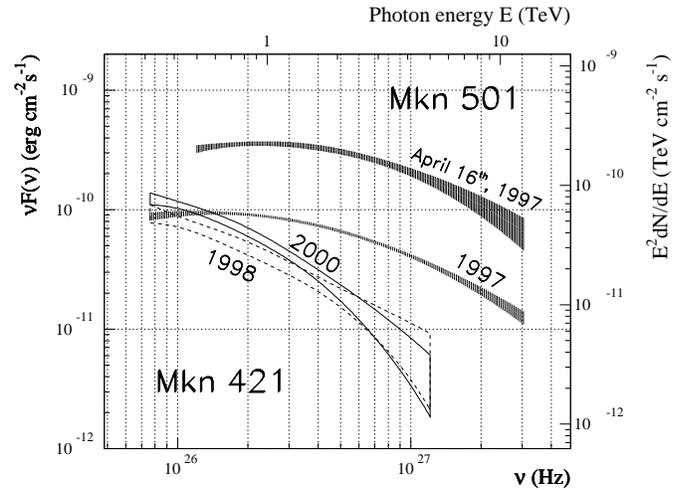,width=\linewidth,clip=
,bbllx=25pt,bblly=45pt,bburx=550pt,bbury=440pt}
\caption[]
{Comparison of the \object{Mkn~421} spectral energy distributions derived in this paper (unfilled areas) with that obtained
by CAT for \object{Mkn~501} in 1997 (hatched areas) (\cite{Djannati99}; \cite{PironThese}).
All areas show the $68$\% confidence level contour given by the likelihood method in the curved shape hypothesis.
}
\label{Fig421vs501}
\end{center}
\end{figure}

The second $\mathrm{TeV}$ blazar, \object{Mkn~501}, underwent a series of intense flares in 1997 observed by all Cherenkov
telescopes in the Northern Hemisphere (\cite{Djannati99} and references therein). 
Fig.~\ref{Fig421vs501} shows two spectral energy distributions (SEDs) of \object{Mkn~501}, as measured by CAT. The
first one is the 1997 time-averaged spectrum and the second one corresponds to the single night of April
16$^\mathrm{th}$, 1997, during which the source reached its highest intensity. The spectral variability observed by CAT 
on the basis of a hardness ratio (\cite{Djannati99}) is illustrated here by the shift in the
peak energy of the SED: $0.56$$\pm$$0.10\:\mathrm{TeV}$ for the average spectrum and $0.94$$\pm$$0.16\:\mathrm{TeV}$ for
the highest flare. Fig.~\ref{Fig421vs501} also shows the SEDs of \object{Mkn~421} for the two observation periods
1998 and 2000: both are obtained with the assumption of a curved spectrum (see Sect.~\ref{SecResSubSecSpectra}).\\

In the framework of the phenomenological unification scheme of blazars established by~\cite{Fossati98}, the broad-band SED of such
a source consists of two components, respectively a low-energy component attributed to synchrotron radiation and a high-energy component
peaking in the $\gamma$-ray range. Peak energies of the preceding components are correlated and \object{Mkn~501}, as observed in 1997, 
appears to be the most extreme blazar, with the synchrotron part of the SED spectrum peaking in the hard X-ray range 
and the $\gamma$-ray component pea\-king around or above $500\:\mathrm{GeV}$.

In the case of \object{Mkn~421}, the CAT 1998 and 2000 data, and the weakness of its detection by EGRET, when interpreted within the preceding
unification scheme, indicate that the peak energy of the high-energy component of the SED lies in the $50\:\mathrm{GeV}$ region, while its
synchrotron peak is known to lie in the UV-soft X-ray band.
A shift of both components towards higher energies during an intense flare, similar to that observed by CAT for \object{Mkn~501} in 1997,
would result in a harder spectrum in the energy region covered by Cherenkov telescopes and thus account for the
trend shown in Fig.~\ref{Fig421cont}. In fact, the spectral hardening of \object{Mkn~421} at $\mathrm{TeV}$ energies during flares, if clearly
proven, should not be a surprise, since a similar effect has been already observed at a few $\mathrm{keV}$ from this source
by the ASCA and Beppo-SAX
satellites (\cite{Takahashi99}; \cite{Malizia00}), and since X-ray and $\gamma$-ray {\it integrated} emissions of \object{Mkn~421} have proven
to be correlated on many occasions (see, e.g., \cite{Maraschi99}; \cite{Takahashi99,Takahashi00}).\\

\section{Conclusions}
\label{SecConcl}
Along with \object{Mkn~501}, \object{Mkn~421} is the most extreme known BL~Lac object. The CAT results presented in this paper confirm the
high variability of this source at $\mathrm{TeV}$ energies, which has been already reported in the past for intense and rapid bursts
(\cite{Gaidos}). Whereas \object{Mkn~501} $\mathrm{TeV}$ va\-ria\-bi\-li\-ty has been observed by CAT from night to night, but never on shorter time-scales,
\object{Mkn~421} showed a fast variation in intensity within one hour during the night from 4 to 5 February,
2000. A simple causality argument implies that the $\gamma$-ray emitting region must be very compact with a size of $\sim$$10$~light-hours
if one takes a typical value of $10$ for the geometric Doppler factor (see, e.g., \cite{Celotti98}), which reduces the
time-scale in the observer frame.

As for the spectral properties, \object{Mkn~421} has proved to be a little less extreme than \object{Mkn~501}. Whereas the latter has
shown a $\gamma$-ray peak lying above the CAT threshold, \object{Mkn~421} exhibited a power-law spectrum in 1998, indicating that current imaging
Cherenkov detectors cover the end part of its spectral energy distribution. However, the 2000 time-averaged spectrum
shows some indication of curvature, which in fact has been also marginally observed by the Whipple Observatory at the time
of the 1995--96 flaring periods (\cite{Krennrich99a}).\\

The present observations of \object{Mkn~421} leave some open questions. Firstly, definite conclusions on a possible spectral
variability still need simultaneous observations of more numerous flares by different Cherenkov telescopes
in order to exclude possible
systematic effects between experiments. Secondly, the interpretation of the multi-wavelength spectra of \object{Mkn~421} still suffers
from the lack of precise {\it simultaneous spectral measurements} of the low and high-energy parts of its SED which could bring accurate
constraints to mo\-dels. For instance, in the Synchrotron Self Compton (SSC) model which is often invoked to explain the SED of extreme
BL~Lac's (\cite{Ghisellini98}), the $\gamma$-ray component is interpreted as the result of the inverse Compton process occurring between
the ultra-relativistic electrons, which emit synchrotron radiation at low energies, and the latter soft photon field itself. This simple
and most natural model predicts a strong correlation between the synchrotron and the $\gamma$-ray bump behaviours. Such a correlation
has been observed many times on \object{Mkn~421}, for example during the coordinated observation campaign in spring 1998, which involved ground-based
Cherenkov imaging telescopes (Whipple, HEGRA, and CAT), and the Beppo-SAX and ASCA X-ray satellites
(\cite{Maraschi99}; \cite{Takahashi99,Takahashi00}). The correlation was only proven in terms of integrated (and not
differential) fluxes due to the lack of statistics. Similarly in 2000, the $\mathrm{TeV}$ flaring behaviour of \object{Mkn~421} was
accompanied by an overall increase of its $\mathrm{keV}$ activity, as for instance con\-ti\-nuous\-ly recorded by the ins\-tru\-ments on board the
Rossi X-Ray Timing Explorer satellite (\cite{PironThese}).

SSC models have been already successfully applied to the Beppo-SAX and CAT data obtained
on \object{Mkn~501} in 1997 (e.g., \cite{PironThese}; \cite{Katarzynski01}), but alternative scenarios still exist also for this source
(e.g., \cite{Rachen99}; \cite{Muecke01}), which consider an e$^-$p plasma with synchrotron photons radiated by electrons at low energies and
$\gamma$-rays emitted at high energies by the products of the proton-induced cascades (from $\pi^0$ decays and from synchrotron
radiation of protons and muons). Thus, in order to address more deeply the crucial problem of the plasma jet content, a richer
sample of flares from both blazars, detected at various wavelengths, would be necessary.\\

\begin{acknowledgements}
The authors wish to thank the French national ins\-ti\-tu\-tions IN2P3/CNRS
and DAPNIA/DSM/CEA for supporting and funding the CAT project.
The CAT telescope was also partly funded by the
Languedoc-Roussillon region and the Ecole Polytechnique. The authors
also wish to thank Electricit\'e de France for making available to them
equipment at the former solar plant ``Th\'emis'' and allowing the 
building of the new telescope and its hangar. They are grateful to the
French and Czech ministries of Foreign Affairs for providing grants
for physicists' travel and accommodation expenses.
L.R. thanks for the financial support granted by the Ministry of
Education of the Czech Republic (Project LN00A006).
\end{acknowledgements}

\appendix
\section{Spectra and light curve extraction}
\subsection{The spectral analysis procedure}
\label{AnnSubSecSp}
Here we give the expression of the likelihood function $\mathcal{L}$ used for spectral reconstruction, using the same notations as in
Sect.~\ref{SecDetSubSecSp}. We start with an assumption on the spectral shape by choosing a given law
$\displaystyle{\left(\frac{\mathrm{d}\phi}{\mathrm{d}E}\right)}^{\mathrm{pred}}$, which includes the set $\{\Lambda\}$ of parameters to fit.
Then, for each 2D-bin $\displaystyle\Delta_{i_z, i_e}$, cor\-res\-pon\-ding to the zenith angle interval
$[\theta^{\mathrm{min}}_{i_z}, \theta^{\mathrm{max}}_{i_z}]$ and to the estimated energy interval 
$[\widetilde{E}^{\mathrm{min}}_{i_e}, \widetilde{E}^{\mathrm{max}}_{i_e}]$, let us define:
\begin{itemize}
\item{$n_{i_z, i_e}$ and $p_{i_z, i_e}$ as the numbers of events passing the se\-lec\-tion cuts in the ON and OFF data, respectively;}
\item{$\beta_{i_z}$ as the normalisation factor between ON and OFF observations, obtained from the contents of the respective
$\alpha$ distributions in the range from $20\degr$ to $120\degr$ (the control region) (see Fig.~\ref{FigAlplot});}
\item{$S_{i_z, i_e}=n_{i_z, i_e}-\beta_{i_z}p_{i_z, i_e}$ as the observed number of $\gamma$-ray events, and
${\delta S}_{i_z, i_e}=\sqrt{n_{i_z, i_e}+\beta_{i_z}^2p_{i_z, i_e}}$ as its error;}
\item{$S_{i_z, i_e}^{\mathrm{pred}}$ as the predicted number of $\gamma$-ray events: if we note $\Upsilon(\theta_{\mathrm z},E_\gamma$$\rightarrow$$\widetilde{E_\gamma})\:{\mathrm d}\widetilde{E_\gamma}$ for
the probability, at fixed zenith angle $\theta_{\mathrm z}$ and {\it real} energy $E_\gamma$, to get an {\it estimated} energy $\widetilde{E_\gamma}$ within the interval
$[\widetilde{E_\gamma}, \widetilde{E_\gamma}$$+$$\mathrm{d}\widetilde{E_\gamma}]$, then
\begin{equation*}
S_{i_z, i_e}^{\mathrm{pred}}=T_{\mathrm{ON}}\:{\int}_{\widetilde{E}^{\mathrm{min}}_{i_e}}^{\widetilde{E}^{\mathrm{max}}_{i_e}}\:
{\mathrm d}\widetilde{E}\:{\int}_{0}^{\infty}\:{\mathrm d}E\:{\left(\frac{\mathrm{d}\phi}{\mathrm{d}E}\right)}^{\mathrm{pred}}\:\times
\end{equation*}
\vspace*{-.5cm}
\begin{equation}
\mathcal{A}_\mathrm{eff}(\overline{\theta_{i_z}},E)\,\Upsilon(\overline{\theta_{i_z}},E\rightarrow\widetilde{E}),
\label{Eq1}
\end{equation}
where $T_{\mathrm{ON}}$ is the total ON-source time of observation, $\mathcal{A}_\mathrm{eff}$ is the effective detection area, and where
$\displaystyle\overline{\theta_{i_z}}$ is defined by
$\displaystyle\cos(\overline{\theta_{i_z}})$$\equiv$$\frac{1}{2}\left[\cos(\theta^{\mathrm{min}}_{i_z})+\cos(\theta^{\mathrm{max}}_{i_z})\right]$;}
\item{$\overline{p_{i_z, i_e}}$ the mean predicted number of events in the OFF data;}
\item{$\overline{n_{i_z, i_e}}=S_{i_z, i_e}^{\mathrm{pred}}+\beta_{i_z}\overline{p_{i_z, i_e}}$ the mean predicted number of events in the ON data.}
\end{itemize}

The observed numbers $n_{i_z, i_e}$ and $p_{i_z, i_e}$ have Poissonian probability distributions
$\mathcal{P}(n_{i_z, i_e})$ and $\mathcal{P}(p_{i_z, i_e})$, respectively,
and the likelihood function is as follows:
\begin{equation}
\displaystyle\mathcal{L}\left(\{\Lambda\}, \{\overline{p_{i_z, i_e}}\}\right)=\prod_{i_z, i_e}\mathcal{P}(n_{i_z, i_e})\mathcal{P}(p_{i_z, i_e}).
\label{Eq2}
\end{equation}

The quantities $\overline{p_{i_z, i_e}}$, which are unknown, can be determined by maximizing the function $\mathcal{L}$:
the relations $\displaystyle\frac{\partial\log(\mathcal{L})}{\partial \overline{p_{i_z, i_e}}}$$\equiv$$0$ lead to the solutions
$\displaystyle\overline{p_{i_z, i_e}}=\frac{1}{2\beta_{i_z}(\beta_{i_z}+1)}\left[a_{i_z, i_e}+\sqrt{a_{i_z, i_e}^2+
4\beta_{i_z}(\beta_{i_z}+1)p_{i_z, i_e}S_{i_z, i_e}^{\mathrm{pred}}}\right]$,
with $\displaystyle a_{i_z, i_e}=\beta_{i_z}(n_{i_z, i_e}+p_{i_z, i_e})-(\beta_{i_z}+1)S_{i_z,i_e}^{\mathrm{pred}}$.\\

Finally, we reinject these expressions in Eq.~(\ref{Eq2}), eliminate the ``constant'' terms which only depend on
$n_{i_z, i_e}$ or $p_{i_z, i_e}$, and get:
\begin{equation*}
\log\left[\mathcal{L}\left(\{\Lambda\}\right)\right]=\sum_{i_z, i_e}\Big[n_{i_z, i_e}\log\left(S_{i_z, i_e}^{\mathrm{pred}}+\beta_{i_z}\overline{p_{i_z,i_e}}\right)
\end{equation*}
\vspace*{-.5cm}
\begin{equation}
+\;\;\;p_{i_z, i_e}\log\left(\overline{p_{i_z, i_e}}\right)
-\left(\beta_{i_z}+1\right)\overline{p_{i_z, i_e}}-S_{i_z, i_e}^{\mathrm{pred}}\Big].
\label{Eq3}
\end{equation}

The $\{\Lambda\}$ parameters, included in this expression through Eq.~(\ref{Eq1}), are the only unknown quantities left: their values
are determined using an iterative procedure of maximization, which leads to the final fitted values and their covariance matrix $V$.

\subsection{Integral flux determination}
\label{AnnSubSecFlux}
When computing an integral flux, one must again take care of the telescope detection threshold
increase with zenith angle $\theta_{\mathrm z}$; thus, let us denote:
\begin{itemize}
\item{$\widetilde{E}^\mathrm{int}[\theta_{\mathrm z}]$ for the unique energy which verifies $\mathcal{A}_\mathrm{eff}(\theta_{\mathrm z},\widetilde{E}^\mathrm{int})\equiv\mathcal{A}_\mathrm{eff}(0\degr,250\:\mathrm{GeV})$;}
\item{$S^\mathrm{int}$ for the total number of $\gamma$-ray events observed within the selection cuts with an {\it estimated}
energy above $\widetilde{E}^\mathrm{int}[\theta_{\mathrm z}]$ ($S^\mathrm{int}$ is determined by a simple ``ON--OFF'' subtraction), and $\delta S^\mathrm{int}$ for its error.}
\end{itemize}
To convert $S^\mathrm{int}$ into the corresponding integral flux $\Phi$ above $250\:\mathrm{GeV}$, we use the following proportionality relation:
\begin{equation*}
\displaystyle\Phi=\frac{S^\mathrm{int}}{S^{\mathrm{int},\;\mathrm{best}}}\times{\int}_{250\:\mathrm{GeV}}^{\infty}\;{\left(\frac{\mathrm{d}\phi}{\mathrm{d}E}\right)}^\mathrm{best}{\mathrm d}E,
\end{equation*}
where $\displaystyle{\left(\frac{\mathrm{d}\phi}{\mathrm{d}E}\right)}^\mathrm{best}$ is the source differential spectrum measured as explained in Sect.~\ref{SecDetSubSecSp},
and $S^{\mathrm{int},\;\mathrm{best}}$ is the number of $\gamma$-ray events predicted from this law above $\widetilde{E}^\mathrm{int}[\theta_{\mathrm z}]$:
\begin{equation*}
\displaystyle S^{\mathrm{int},\;\mathrm{best}}=T_{\mathrm{ON}}\:{\int}_{\widetilde{E}^\mathrm{int}[\theta_{\mathrm z}]}^{\infty}\:{\mathrm d}\widetilde{E}\:
{\int}_{0}^{\infty}\:{\mathrm d}E\:{\left(\frac{\mathrm{d}\phi}{\mathrm{d}E}\right)}^\mathrm{best}\:\times
\end{equation*}
\vspace*{-.5cm}
\begin{equation*}
\mathcal{A}_\mathrm{eff}(\theta_{\mathrm z},E)\:\Upsilon(\theta_{\mathrm z},E\rightarrow\widetilde{E}).
\end{equation*}
Finally, the error on $\Phi$ is simply estimated as $\displaystyle\delta\Phi=\frac{\delta S^\mathrm{int}}{S^\mathrm{int}}\times\Phi$.

\section{More on \object{Mkn~421} spectra}
\label{AnnSecSp421}
\newcommand{\zzd}[9]{
\multicolumn{3}{c}{$\zzs{T_{\mathrm{ON}}=#1\:\mathrm{h}}$}&\multicolumn{2}{c|}{$\zzs{\theta_{\mathrm z}\in[#2^\circ;#3^\circ]}$}&
\multicolumn{3}{c}{$\zzs{T_{\mathrm{ON}}=#4\:\mathrm{h}}$}&\multicolumn{2}{c|}{$\zzs{\theta_{\mathrm z}\in[#5^\circ;#6^\circ]}$}&
\multicolumn{3}{c}{$\zzs{T_{\mathrm{ON}}=#7\:\mathrm{h}}$}&\multicolumn{2}{c}{$\zzs{\theta_{\mathrm z}\in[#8^\circ;#9^\circ]}$}
}
\newcommand{\zze}[9]{
\multicolumn{3}{c}{$\zzs{T_{\mathrm{OFF}}=#1\:\mathrm{h}}$}&\multicolumn{2}{c|}{$\zzs{S_\gamma=#2\pm#3}$}&
\multicolumn{3}{c}{$\zzs{T_{\mathrm{OFF}}=#4\:\mathrm{h}}$}&\multicolumn{2}{c|}{$\zzs{S_\gamma=#5\pm#6}$}&
\multicolumn{3}{c}{$\zzs{T_{\mathrm{OFF}}=#7\:\mathrm{h}}$}&\multicolumn{2}{c}{$\zzs{S_\gamma=#8\pm#9}$}
}
\newcommand{\zzf}{$\zzs{n_{i_e}}$&$\zzs{p_{i_e}}$&$\zzs{S_{i_e}}$&$\zzs{\delta S_{i_e}}$&$\zzs{{N_\sigma}_{i_e}}$}
\newcommand{\zzg}[5]{$\zzs{#1}$&$\zzs{#2}$&$\zzs{#3}$&$\zzs{#4}$&$\zzs{#5}$}

\begin{table*}[t]
\begin{center}
\caption[]{\object{Mkn~421} available statistics.}
\begin{tabular}{c|rrrrr|rrrrr|rrrrr}
\hline
&\multicolumn{5}{c|}{1998}&\multicolumn{5}{c|}{2000 (except 4--5/02)}&\multicolumn{5}{c}{4--5/02/2000}\\
&\zzd{6.2}{0}{28}{8.4}{0}{28}{3.5}{0}{28}\\
&\zze{9.5}{735}{57}{16.7}{1424}{71}{17.6}{609}{40}\\
\hline
$\zzs{\Delta_{i_e}}$&\zzf&\zzf&\zzf\\
\hline
$\zzs{0.3}-\zzs{0.5}$&\zzg{615}{445}{170}{29}{5.8}&\zzg{1351}{993}{358}{44}{ 8.2}&\zzg{464}{318}{146}{23}{ 6.4}\\
$\zzs{0.5}-\zzs{0.8}$&\zzg{790}{536}{254}{35}{7.3}&\zzg{1267}{731}{536}{41}{13.1}&\zzg{561}{300}{261}{25}{10.4}\\
$\zzs{0.8}-\zzs{1.3}$&\zzg{452}{301}{151}{26}{5.7}&\zzg{ 690}{411}{279}{30}{ 9.2}&\zzg{300}{172}{128}{18}{ 7.0}\\
$\zzs{1.3}-\zzs{2.0}$&\zzg{229}{123}{107}{18}{5.9}&\zzg{ 335}{158}{175}{21}{ 8.6}&\zzg{138}{ 64}{ 74}{12}{ 6.0}\\
$\zzs{2.0}-\zzs{5.0}$&\zzg{119}{ 66}{ 53}{13}{4.0}&\zzg{ 165}{ 89}{ 76}{15}{ 5.1}&\zzg{-}{-}{-}{-}{-}\\
\hline
\end{tabular}
\label{TabSpStat}
\end{center}
\end{table*}

\newcommand{\zzh}[9]{$\zzs{#1}$&$\zzs{#2}$&$\zzs{#3}$&$\zzs{#4}$&$\zzs{#5}$&$\zzs{#6}$&$\zzs{#7}$&$\zzs{#8}$&$\zzs{#9}$\\}

\begin{table*}[t]
\begin{center}
\caption[]{Covariance matrices elements of \object{Mkn~421} spectra.}
\begin{tabular}{lccccccccc}
\hline
\footnotesize{Period}&$\zzs{V_{\phi\phi}^\mathrm{pl}}$&$\zzs{V_{\phi\gamma}^\mathrm{pl}}$&$\zzs{V_{\gamma\gamma}^\mathrm{pl}}$&
$\zzs{V_{\phi\phi}^\mathrm{cs}}$&$\zzs{V_{\phi\gamma}^\mathrm{cs}}$&$\zzs{V_{\phi\beta}^\mathrm{cs}}$&$\zzs{V_{\gamma\gamma}^\mathrm{cs}}$&$\zzs{V_{\gamma\beta}^\mathrm{cs}}$&$\zzs{V_{\beta\beta}^\mathrm{cs}}$\\
\hline
$\zzs{1998}$&\zzh{3.97\times 10^{-2}}{-1.24\times 10^{-2}}{1.44\times 10^{-2}}
{1.03\times 10^{-1}}{-1.10\times 10^{-2}}{1.13\times 10^{-1}}{2.35\times 10^{-2}}{3.28\times 10^{-2}}{2.35\times 10^{-1}}
$\zzs{2000}$&\zzh{3.19\times 10^{-2}}{-8.65\times 10^{-3}}{6.51\times 10^{-3}}
{7.80\times 10^{-2}}{-8.36\times 10^{-4}}{6.45\times 10^{-2}}{1.68\times 10^{-2}}{2.91\times 10^{-2}}{1.36\times 10^{-1}}
$\zzs{4-5/02/2000}$&\zzh{1.12\times 10^{-1}}{-3.84\times 10^{-2}}{2.16\times 10^{-2}}
{1.38\times 10^{-1}}{-3.45\times 10^{-2}}{3.63\times 10^{-2}}{1.53\times 10^{-1}}{2.86\times 10^{-1}}{6.57\times 10^{-1}}
\hline
\end{tabular}
\label{TabSpCov}
\end{center}
\end{table*}

\subsection{Available statistics}
\label{AnnSecSp421SubSecStat}
The statistics used for extracting the \object{Mkn~421} spectra are given in Table~\ref{TabSpStat}. With the same notations
used in this paper, we write:
\begin{itemize}
\item{the data zenith angle amplitude, and the ON and OFF-source observation durations, $T_{\mathrm{ON}}$ and $T_{\mathrm{OFF}}$;}
\item{for each of the $n_e$ energy intervals
$\Delta_{i_e}\equiv[\widetilde{E}^{\mathrm{min}}_{i_e}, \widetilde{E}^{\mathrm{max}}_{i_e}]$ (in $\mathrm{TeV}$):}
\begin{description}
\item{$\bullet$ the numbers of events passing the selection cuts in the ON and OFF data,\\
$n_{i_e}=\sum_{i_z}n_{i_z,i_e}$ and
$p_{i_e}=\sum_{i_z}\beta_{i_z}p_{i_z,i_e}$;}
\item{$\bullet$ the observed number of $\gamma$ events, $S_{i_e}=n_{i_e}-p_{i_e}$, and its error,
$\delta S_{i_e}={\left({\sum_{i_z}n_{i_z,i_e}+{\beta_{i_z}}^2p_{i_z,i_e}}\right)}^{1/2}$;}
\item{$\bullet$ the statistical significance ${N_\sigma}_{i_e}=S_{i_e}/\delta S_{i_e}$;}
\end{description}
\item{the observed total number of $\gamma$ events, $S_\gamma=\sum_{i_e}S_{i_e}$, and its error,
$\delta S_\gamma={\left({\sum_{i_e}{\delta S_{i_e}}^2}\right)}^{1/2}$.}
\end{itemize}

\subsection{Covariance matrices}
\label{AnnSecSp421SubSecCov}
Let $V^\mathrm{pl}$ and $V^\mathrm{cs}$ be the two covariance matrices of spectral parameters for the $\mathcal{H}^\mathrm{pl}$ and $\mathcal{H}^\mathrm{cs}$ hypotheses
(see Sect.~\ref{SecDetSubSecSp}), res\-pec\-ti\-ve\-ly:
\begin{displaymath}
V^\mathrm{pl}=\left(\begin{array}{cc}
V_{\phi\phi}^\mathrm{pl}&V_{\phi\gamma}^\mathrm{pl}\\
V_{\phi\gamma}^\mathrm{pl}&V_{\gamma\gamma}^\mathrm{pl}\\
\end{array}\right)
\;\;\mathrm{and}\;\;
V^\mathrm{cs}=\left(\begin{array}{ccc}
V_{\phi\phi}^\mathrm{cs}&V_{\phi\gamma}^\mathrm{cs}&V_{\phi\beta}^\mathrm{cs}\\
V_{\phi\gamma}^\mathrm{cs}&V_{\gamma\gamma}^\mathrm{cs}&V_{\gamma\beta}^\mathrm{cs}\\
V_{\phi\beta}^\mathrm{cs}&V_{\gamma\beta}^\mathrm{cs}&V_{\beta\beta}^\mathrm{cs}\\
\end{array}\right).
\end{displaymath}
The values of these matrices elements are given in Table~\ref{TabSpCov} for each spectrum of \object{Mkn~421} obtained in this paper.\\

In the power-law hypothesis, the energy $E_\mathrm{d}$ at which the values of $\phi_0^\mathrm{pl}$ et $\gamma^\mathrm{pl}$ are decorrelated writes:
\begin{displaymath}
E_\mathrm{d}=\exp\left[\frac{V_{\phi\gamma}^\mathrm{pl}}{\phi_0^\mathrm{pl}V_{\gamma\gamma}^\mathrm{pl}}\right]\:\mathrm{TeV}.
\end{displaymath}
At this decorrelation energy, the width of the hatched area re\-pre\-sen\-ting the $68$\% confidence level contour, is minimum: for
instance, we find $E_\mathrm{d}$$=$$690\:\mathrm{GeV}$ for the 1998 time-averaged spectrum of \object{Mkn~421} shown in Sect.~\ref{SecResSubSecSpectra}.

In the curved shape hypothesis, the energy-dependent exponent
$\gamma^\mathrm{cs}_l(E_{\mathrm{TeV}})=\gamma^\mathrm{cs}+\beta^\mathrm{cs}\log_{10}E_{\mathrm{TeV}}$ has a minimal error at the energy
$\displaystyle E^\mathrm{cs}_0=10^{\frac{-V_{\gamma\beta}^\mathrm{cs}}{V_{\beta\beta}^\mathrm{cs}}}\:\mathrm{TeV}$; the corresponding value of $\gamma^\mathrm{cs}_l$ writes
\begin{displaymath}
\gamma^\mathrm{cs}_0\equiv\gamma^\mathrm{cs}_l(E^\mathrm{cs}_0)=\gamma^\mathrm{cs}-\beta^\mathrm{cs}\frac{V_{\gamma\beta}^\mathrm{cs}}{V_{\beta\beta}^\mathrm{cs}},
\end{displaymath}
and its error
\begin{displaymath}
\delta\gamma^\mathrm{cs}_0\equiv\delta\gamma^\mathrm{cs}_l(E^\mathrm{cs}_0)=\sqrt{V_{\gamma\gamma}^\mathrm{cs}-\frac{{V_{\gamma\beta}^\mathrm{cs}}^2}{V_{\beta\beta}^\mathrm{cs}}}.
\end{displaymath}

\end{document}